\documentstyle[aps,preprint,epsfig,floats]{revtex}

\newcommand{\beq}{\begin{equation}}
\newcommand{\eeq}{\end{equation}}
\newcommand{\bdis}{\begin{displaymath}}
\newcommand{\edis}{\end{displaymath}}
\newcommand{\bea}{\begin{eqnarray}}
\newcommand{\eea}{\end{eqnarray}}
\newcommand{\barr}{\begin{array}}
\newcommand{\earr}{\end{array}}

\newcommand{\equ}[1]{(\protect\ref{#1})}

\newcommand{\xpe}{x_\perp}
\newcommand{\vxpe}{\vec{x}_\perp}
\newcommand{\xpa}{x_\parallel}

\begin{document}
\draft
\tightenlines

\title{Critical behavior and conservation in directed sandpiles}
\author{Romualdo Pastor-Satorras and Alessandro Vespignani}

\address{
The Abdus Salam International Centre for Theoretical Physics (ICTP), 
P.O. Box 586, 34100 Trieste, Italy}
 
\date{\today}

\maketitle

\begin{abstract}
  We perform large-scale simulations of directed sandpile models with
  both deterministic and stochastic toppling rules. Our results show
  the existence of two distinct universality classes. We also provide
  numerical simulations of directed models in the presence of bulk
  dissipation. The numerical results indicate that the way in which
  dissipation is implemented is irrelevant for the determination of
  the critical behavior. The analysis of the self-affine properties of
  avalanches shows the existence of a subset of super-universal
  exponents, whose value is independent of the universality class.
  This feature is accounted for by means of a phenomenological
  description of the energy balance condition in these models.
\end{abstract}



\section{Introduction}

Sandpile cellular automata are the most famous example of
self-organized critical (SOC) behavior \cite{btw1,btw2,jensen98}.
Under an external drive consisting of a slow addition of sand (energy)
grains and the action of dissipation through the loss of energy on the
lattice boundaries, these models reach a stationary steady state.  In
the limit of infinitesimal driving and dissipation (this last achieved
in the thermodynamic limit), the stationary state of sandpile models
exhibits diverging response functions associated to a characteristic
avalanche dynamics.  This is the hallmark of a critical behavior that
has attracted an enormous amount of interest as a plausible
explanation of the avalanche-like critical behavior empirically
observed in many natural systems \cite{jensen98}.

Sandpile models have been at the center of an intense research
activity made of both analytical studies and numerical simulations.
Despite the simple definition of these automata, it turns out that
their full analytical understanding is a very problematic task
\cite{dhar99}.  As a further complication, also the numerical
inspection of these models results to be particularly difficult.  For
example, the precise identification of universality classes has
resisted for many years even the most careful numerical analysis, and
only recent results have partially settled this problem
\cite{milshtein98,granada,lubeck00,vdmz99}.  On the other hand, these
refined analyses have pointed out that several sandpile models do not
follow the simple finite size scaling (FSS) form usually adopted in
the description of critical behavior \cite{cardy88}.  For instance,
the more sophisticated multiscaling approach
\cite{kad89,men,tebaldi99} seems to be required for a full description
of the scaling properties of the original Bak, Tang, and Wiesenfeld
(BTW) model \cite{btw1,btw2}.

Many sandpile features have been underlined as the possible origin of
these scaling anomalies. The deterministic dynamical rules of the BTW
model induce nonergodic effects \cite{vdmz99}, that are certainly
missing in stochastic models, such as the Manna model
\cite{manna91b,dhar99}, which shows a perfect FSS behavior, even for
moderate system sizes. A further complication of sandpile automata
stems from the peculiar role of the boundary dissipation, that makes
the lattice size scaling entangled with the system dynamics. In such
cases, the thermodynamic limit is essential for the dissipative
dynamics of large avalanches.  A clear understanding of the interplay
between dissipation and size scaling has not yet been achieved and it
has been recently the subject of several studies \cite{men,drossel00}.

In this paper we address some of the aforementioned problems in the
case of directed sandpile models
\cite{kad89,dhar89,tadic97,hasty98,pv99jpa}. In this case Dhar and
Ramaswamy obtained an exact solution for the Abelian deterministic
directed sandpile (DDS)\cite{dhar89}, that can be used as a milestone
to check the numerical simulation analysis.  Directed sandpiles thus
become an interesting test field to study how the critical behavior is
affected by the introduction of stochastic elements and dissipation.
We perform large scale numerical simulations of two directed sandpile
automata: the deterministic directed sandpile model \cite{dhar89} and
the stochastic directed sandpile model \cite{pv99jpa}.  We study both
models in the case of boundary and bulk dissipation
\cite{manna90b,christensen93b,chessa98,manna99}. We find, in agreement
with the results in Ref.~\cite{pv99jpa}, that the models define two
different universality classes.  In addition we show that the
universality class of the models does not depend on the way in which
dissipation is implemented. Finally we analyze the properties of
anisotropic models in which the dynamics is not fully directed
\cite{kad89,tsuchiya99}. In this case we observe that on large scales
the critical behavior is the same of that of fully directed models.
Results for the stochastic model are compared with a recent
theoretical approach by Paczuski and Bassler\cite{pacz00}, that
provides values for the critical exponents in perfect agreement with
numerical simulations. These results are also recovered in
Ref.~\cite{kloster00}.

The numerical analysis also points out that some critical exponent
values, such as the correlation length exponents or the affinity
exponent (to be defined later on), are independent of the particular
universality classes and common to all models considered.  In order to
explain this numerical evidence, we provide a phenomenological
characterization of directed sandpiles based on the basic symmetries
introduced by the conserved dynamics of these automata. Following
balance of energy arguments inspired in Refs.
\cite{vz98,dvz98,vdmz98}, we derive a series of results and
predictions on the value of critical exponents which are a
straightforward consequence of conservation.  These general results
can be considered as super-universal, because they characterize the
critical behavior of all directed sandpiles with local dynamical
rules, independently on the specific universality class.  The results
presented here provide a general picture of directed models and the
role of boundary and bulk dissipation in the process of
self-organization.

The paper is arranged as follows. In Sec.~II we introduce and define
the various directed models considered.  Secs.~III and IV present and
discuss, from the standpoint of universality, the numerical results
for directed models with boundary and bulk dissipation. In Sec.~V we
introduce anisotropic models, and present the numerical results
obtained, in comparison with those of directed models. Sec.~VI is
devoted to an analytical approach based on the conservation of energy.
Finally, in Sec.~VII we draw our conclusions and perspectives.

\section{Directed Models}
\label{sec:models}

Sandpile models are usually defined on a $d$-dimensional hyper-cubic
lattice of size $L$. To each node of the lattice is assigned an
integer variable $z_i$, called ``energy''. Energy is added to the
system uniformly at randomly chosen sites ($z_i \to z_i + 1$). When a
site becomes active, that is, when its energy becomes larger than or
equal to a certain threshold $z_c$, it topples.  A toppling site loses
an energy $z_c$, that is distributed among its neighbors according to
a certain set of rules. The neighbors that receive energy can become
active and topple on their turn, thus generating an avalanche.  The
slow driving condition is effectively imposed by stopping the random
energy addition during the avalanche spreading. This means that the
driving time scale is infinitely large with respect to the toppling
characteristic time scale.

The models we consider in this Section are directed, in the sense that
the energy is always transported along a preferred fixed direction.
We denote this preferred direction by the coordinate $\xpa$, whose
positive direction is usually defined as ``downwards''. The transverse
direction (subspace of dimension $d-1$ perpendicular to $\xpa$) will
be denoted by $\vxpe$.

\begin{figure}[t]
  \centerline{\epsfig{file=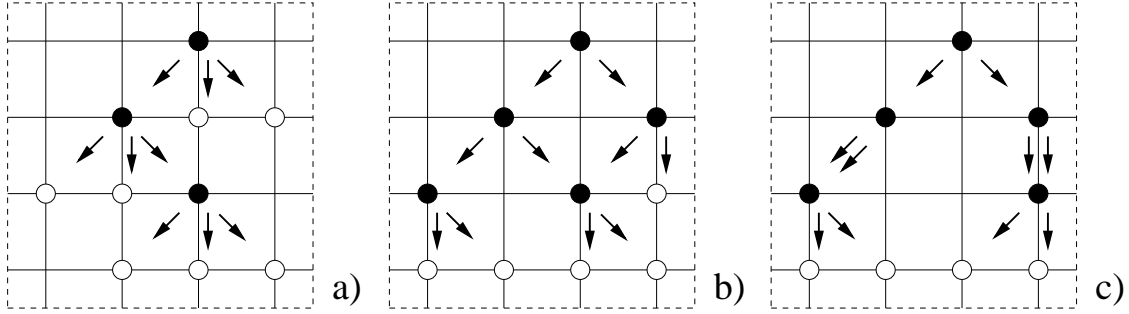, width=15cm}}
  \vspace*{0.5cm}
  \caption{Toppling rules in $d=2$ for directed sandpiles. Filled
    circles represent active (toppling) sites; empty circles are
    stable sites. In the deterministic model (a) an active site sends
    one grain to each of its three neighbors on the next downwards
    row. In the stochastic models, exclusive (b) and nonexclusive (c),
    one grain is sent to two randomly chosen downwards neighbors.}
  \label{fig:rules}
\end{figure}

The toppling rules of the models define two main classes:

(i) {\em Deterministic directed sandpile} (DDS): In $d$ dimensions,
the threshold is set to $z_c=2d+1$. When a site in a given hyper-plane
$\xpa$ topples, it sends {\em deterministically} one grain of energy
to each one of its nearest and next-nearest neighbors on the
hyper-plane $\xpa+1$ (see Fig.~\ref{fig:rules}a)). Our definition is
somewhat different from the original model of Dhar and Ramaswamy
\cite{dhar89}, in both the driving and the orientation of the lattice.
Both models, however, are expected to share the same universality
class, being deterministic and directed. Numerical simulations confirm
indeed this point\cite{pv99jpa}.

(ii) {\em Stochastic directed sandpile} (SDS): In this case, the
threshold is $z_c=2$, independently of the dimensionality of the
lattice. When a site in the hyper-plane $\xpa$ topples, it sends two
grains of energy to two sites, {\em randomly chosen} among its $2d-1$
nearest and next-nearest neighbors on the hyper-plane $\xpa+1$.  The
toppling rules of this model can be defined {\em exclusive} if the two
energy grains are always distributed on different sites,
Fig.~\ref{fig:rules}b). On the other hand, the model can be defined
{\em non-exclusive} if the dynamics allows the transfer of two energy
grains onto the same site, Fig.~\ref{fig:rules}c). We therefore report
simulations on the exclusive stochastic directed sandpile (ESDS) and
on the non-exclusive stochastic directed sandpile (NESDS).  In spite
of the stochastic nature of these models, we must bear in mind that
they are nevertheless Abelian \cite{dhar99}. The discussion therefore
focuses on the difference between stochastic and deterministic models.

Once the toppling rules have been determined, the models are finally
defined by specifying the dissipation mechanism. For systems with
boundary dissipation, we impose periodic boundary conditions in the
transverse directions $\vxpe$ and open at the hyper-plane $\xpa=L$. In
this way, the models are locally conserved; energy can only leave the
system at the bottom of the lattice. In models with bulk dissipation,
we impose periodic boundary conditions in both the $\xpa$ and $\vxpe$
directions. Dissipation is implemented by allowing a toppling site to
loose an energy $z_c$ without transferring it with probability
$p$\cite{vz98,chessa98}.  This means that, on average, an energy
$\epsilon=z_c p$ is dissipated in each toppling. In the limit
$\epsilon\to 0$, the system shows critical behavior \cite{vz98}.

In the stationary state we can define the probability that the
addition of a single energy grain is followed by an avalanche of
toppling events. Avalanches are then characterized by the total number
of topplings $s$ and the time duration $t$. In the limit of
infinitesimal driving (slow driving condition) the system shows
scaling behavior and the probability distributions of these quantities
follow the finite-size scaling (FSS) forms:
\begin{eqnarray}
  P(s)&=&s^{-\tau_s}{\cal G}(s/s_c)\ ,\label{ps} \\
  P(t)&=&t^{-\tau_t}{\cal F}(t/t_c)\ ,\label{pt}
\end{eqnarray}
where $s_c$ and $t_c$ are the characteristic size and time,
respectively. The exponents $\tau_s$ and $\tau_t$ characterize the
critical behavior and define the universality classes to which the
models belong.  In the critical region the characteristic time and
size are determined only by the system size $L$ or the dissipation
$\epsilon$, in the case of boundary and bulk dissipation,
respectively.  In directed models, the {\em affinity exponent} $\zeta$
is of particular importance; it relates the avalanche characteristic
lengths in the perpendicular direction, $\xi_\perp$, and in the
parallel direction, $\xi_\parallel$, through the relation
$\xi_\perp\sim\xi_\parallel^\zeta$. This exponent characterizes the
degree of anisotropy due to the preferential direction present in the
transport of the energy. In other words, it expresses the self-affine
properties in the scaling of avalanches.  A general result concerns
the average avalanche size $\left<s\right>$, that also scales linearly
with $L$ \cite{kad89,dhar89,tsuchiya99}: a new injected grain of
energy has to travel, on average, a distance of order $L$ before
reaching the boundary. In the stationary state, to each energy grain
drop must correspond, on average, an energy grain flowing out of the
system.  This implies that the average avalanche size corresponds to
the average number of topplings needed for a grain to reach the
boundary; i.e., $\left<s\right>\sim L$.  The same result can be
exactly obtained by inspecting the conservation symmetry of the model
as we shall see in Sec. \ref{sec:conservation}.

For the DDS, the exact analytical solution in $d=2$ yields the
exponents $\tau_s=4/3$ and $\tau_t=D=3/2$ \cite{dhar89}.  The upper
critical dimension is found to be $d_c=3$, and it is also possible to
find exactly the logarithmic corrections to scaling
\cite{dhar89,lubeck98}.  The introduction of stochastic ingredients in
the toppling dynamics of directed sandpiles has been studied recently
in a model that randomly stores energy on each toppling
\cite{tadic97}. This model is strictly related to directed percolation
and defines a universality class ``per se''.  In our case
stochasticity affects only the partition of energy during topplings,
and there is {\em a priori} no obvious relations between  the
critical behavior of these models.

\section{Numerical simulations with boundary dissipation}

In this Section we report results from computer simulations of
deterministic and stochastic directed sandpiles, performed with
boundary dissipation. The system sizes considered range from $L=100$
to $L=6400$. The statistical distribution functions have been computed
averaging over $10^7$ nonzero avalanches.

In the case of boundary dissipation, the lattice size $L$ is the only
characteristic length present in the system. Approaching the
thermodynamic limit ($L\to\infty$), the avalanche characteristic size
and time in Eqs.~\ref{ps} and~\ref{pt} diverge as $s_c\sim L^D$ and
$t_c\sim L^z$, respectively.  The exponent $D$ defines the fractal
dimension of the avalanche cluster and $z$ is the usual dynamic
critical exponent. The directed nature of the model introduces a
drastic simplification, since it imposes $z=1$.  In order to compute
the different exponents characterizing the dynamics of the avalanches,
we have performed the moment analysis of the distributions, in analogy
to the method developed by De Menech {\em et al.}
\cite{men,tebaldi99}.  We define the $q$-th moment of the avalanche
size distribution on a lattice of size $L$ as $\left< s^q \right>_L =
\int ds \, s^q \, P(s)$. If the FSS hypothesis \equ{ps} is valid in
the asymptotic limit of large $s$, then the $q$-th moment has the
following dependence on system size:
\begin{equation}
  \left< s^q \right>_L = L^{D(q+1-\tau_s)} \int dy \, y^{(q-\tau_s)}\, 
  {\cal   G}(y) \sim L^{\sigma_s(q)}.
  \label{eq:moments}
\end{equation}
The exponent $\sigma_s(q)=D(q+1-\tau_s)$ is computed as the slope of
the log-log plot of $\left< s^q \right>_L$ as a function of $L$. For
large enough values of $q$ (i.e., away from the region where the
integral in \equ{eq:moments} is dominated by its lower cut-off), one
can compute the fractal dimension $D$ as the slope of $\sigma_s(q)$ as
a function of $q$: $D=\partial \sigma_s(q)/ \partial q$. On the other
hand, as we have argued in the previous Section, the first moment must
scale linearly with $L$, which imposes $\sigma_s(1)=1$.  Once $D$ is
known we can estimate $\tau_s$ using the relation
$\sigma_s(1)=D(2-\tau_s)$

Along the same lines we can obtain the moments of the avalanche time
distribution. In this case, $\langle t^q\rangle_L\sim
L^{\sigma_t(q)}$, with $\partial\sigma_t(q)/\partial q=z$. Analogous
considerations for small $q$ apply also for the time moment analysis.
Here, an estimate of the asymptotic convergence of the numerical
results is provided by the constraint $z=1$, that must hold for large
enough sizes. Then, the $\tau_t$ exponent can be found using the
scaling relation $(2-\tau_t)=\sigma_t(1)$.

Once the exponents have been estimated numerically, we can check the
accuracy of the moment analysis' predictions using the FSS hypothesis.
If the FSS hypothesis of Eq.s~(\ref{ps},\ref{pt}) is correct, then the
plots of the distributions, under the rescaling $s\to s/L^D$ and
$P(s)\to P(s)L^{D\tau_s}$ and correspondingly $t\to t/L^z$ and
$P(t)\to P(t)L^{z\tau_t}$, should collapse onto the same universal
function, for different values of $L$.

\begin{table}[b]
\begin{tabular}{lccccc}
  Model & $\tau_s$   & $D$   & $\tau_t$   & $z$ & $\zeta$\\
\hline
DR & $4/3$ & $3/2$ & $3/2$ & $1$  & $1/2$ \\
DDS  &  $1.34(1)$ & $1.51(1)$ & $1.51(1)$ & $1.00(1)$ & $0.50(1)$\\
ESDS  &  $1.43(1)$ & $1.74(1)$ & $1.71(3)$ & $0.99(1)$ & $0.51(1)$\\
NESDS  &  $1.43(1)$ & $1.75(1)$ & $1.74(4)$ & $0.99(1)$ & $0.51(1)$
\end{tabular}
\caption{Critical exponents for directed sandpiles with boundary
  dissipation in $d=2$. DR: Dhar and Ramaswamy's exact solution; DDS,
  deterministic directed model; ESDS and NESDS, stochastic directed
  models. Figures in parenthesis denote statistical uncertainties.} 
\label{tableL}
\end{table}

\begin{figure}[t]
  \centerline{\epsfig{file=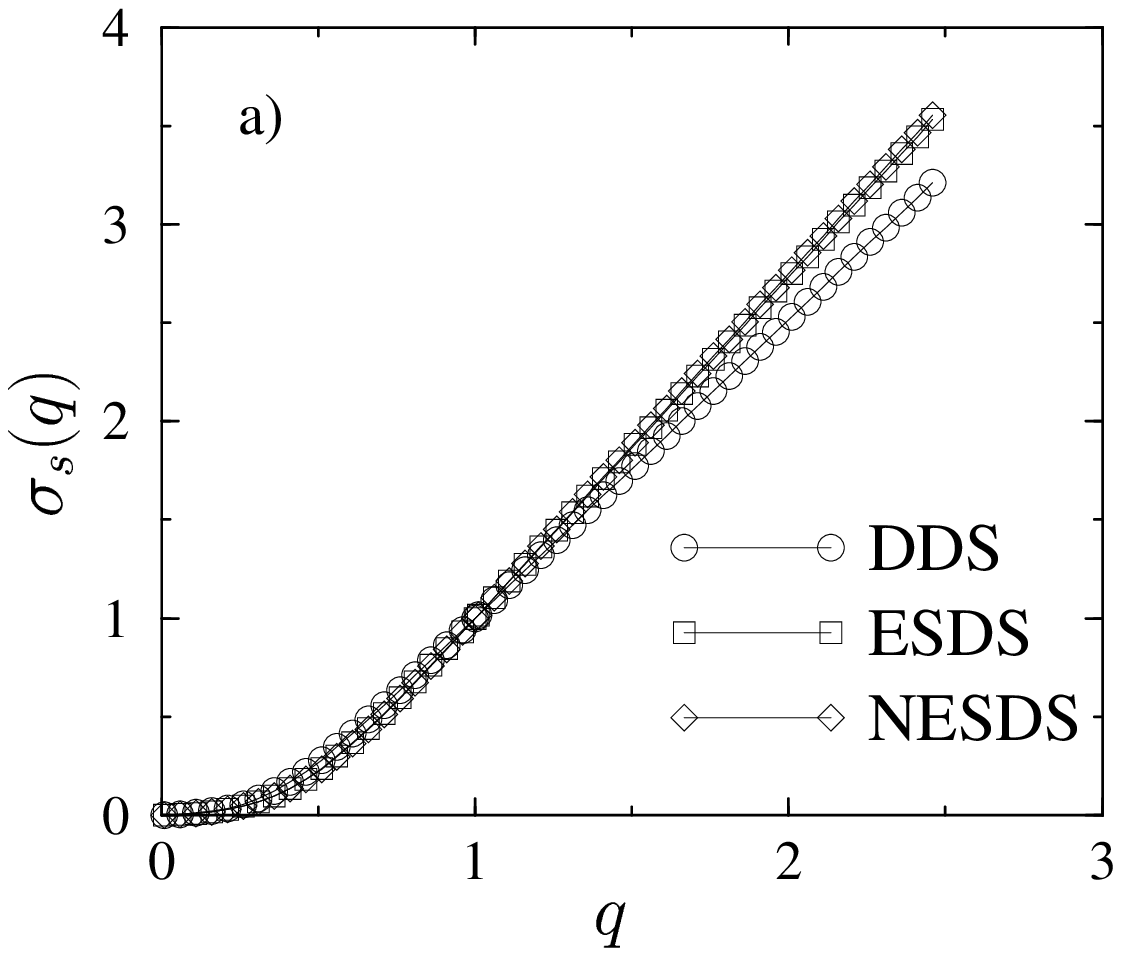, width=7cm} 
    \epsfig{file=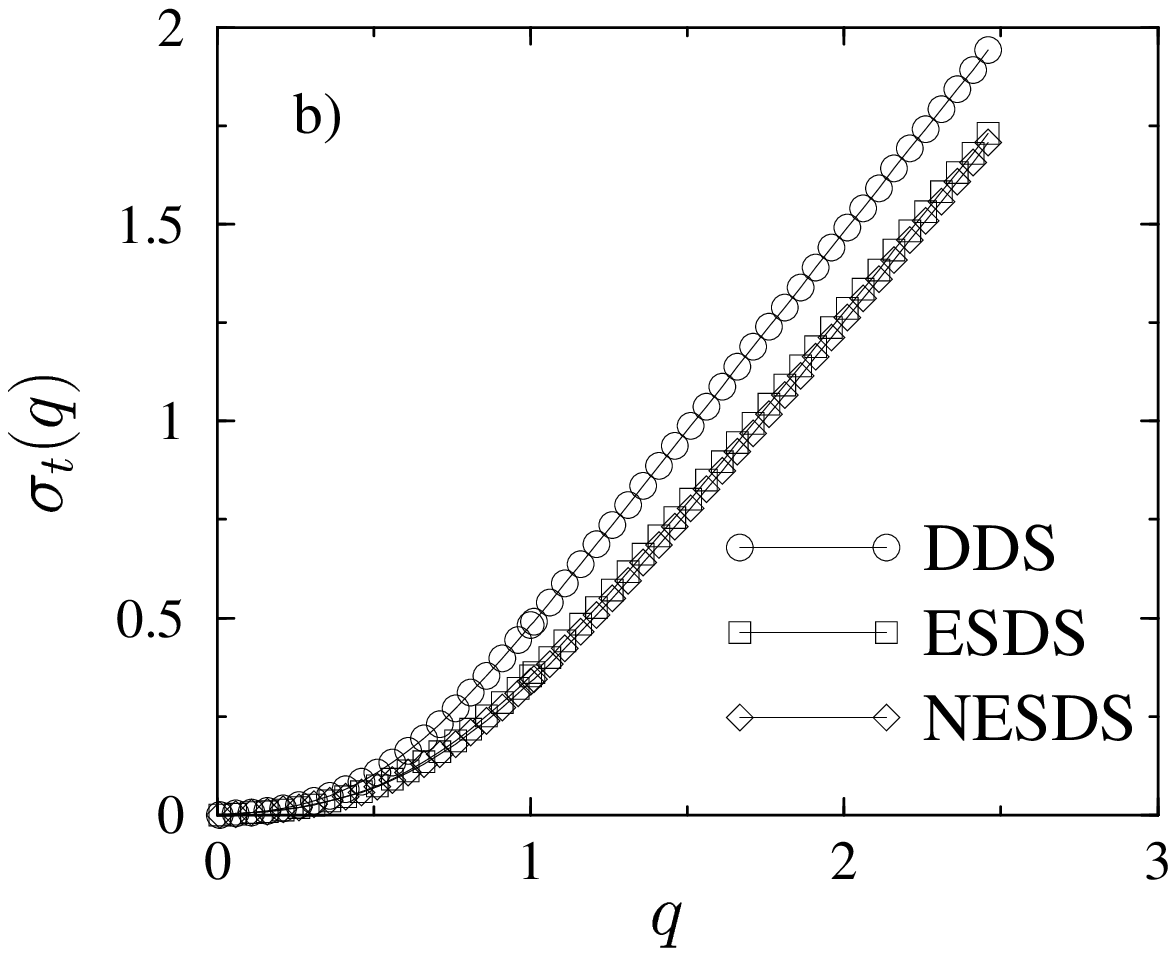, width=7cm}}
  \vspace*{0.25cm}
  \caption{Plot of (a) $\sigma_s(q)$ and (b) $\sigma_t(q)$ for the
    $d=2$ models DDS, ESDS, and NESDS with boundary dissipation.}
  \label{fig:momentsL}
\end{figure}

\begin{figure}[t]
  \centerline{\epsfig{file=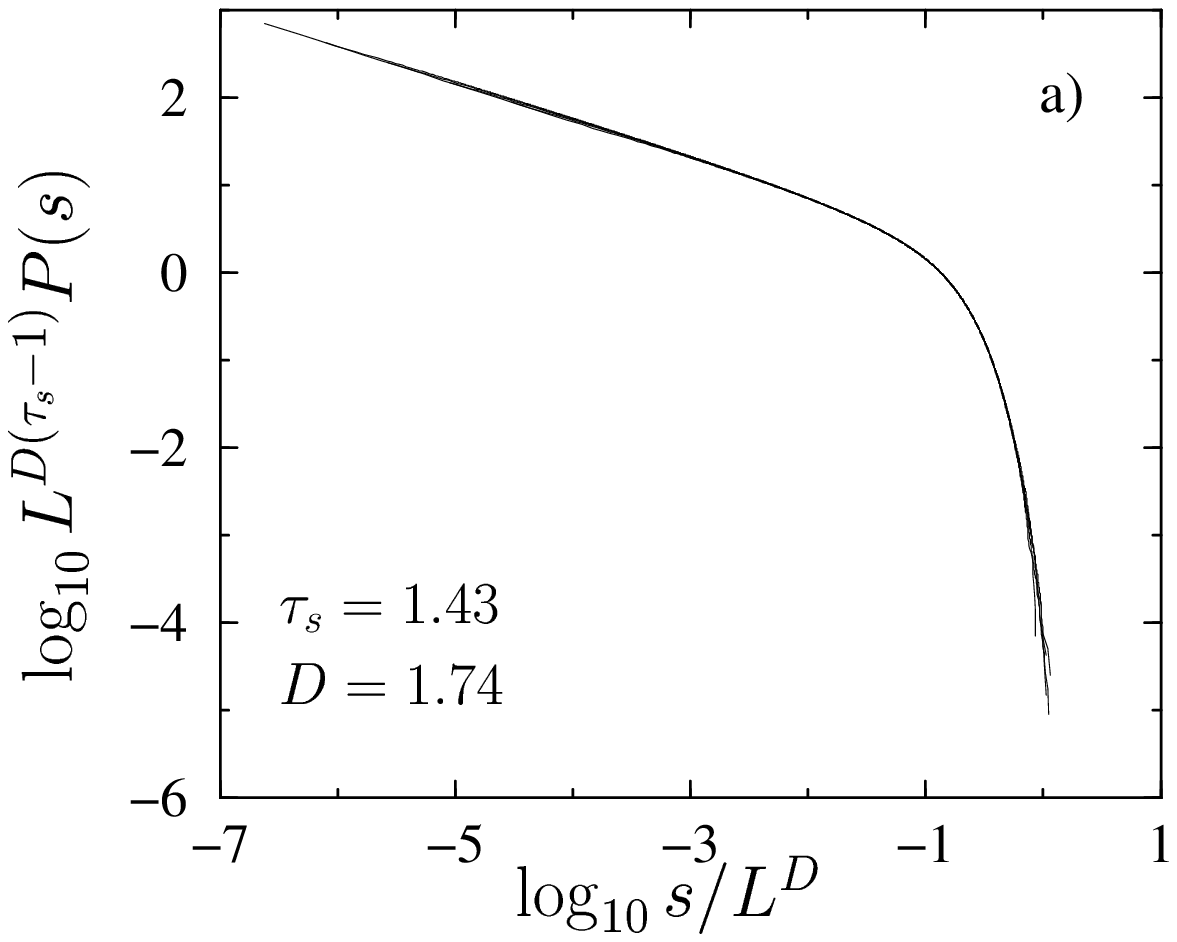, width=7cm} 
    \epsfig{file=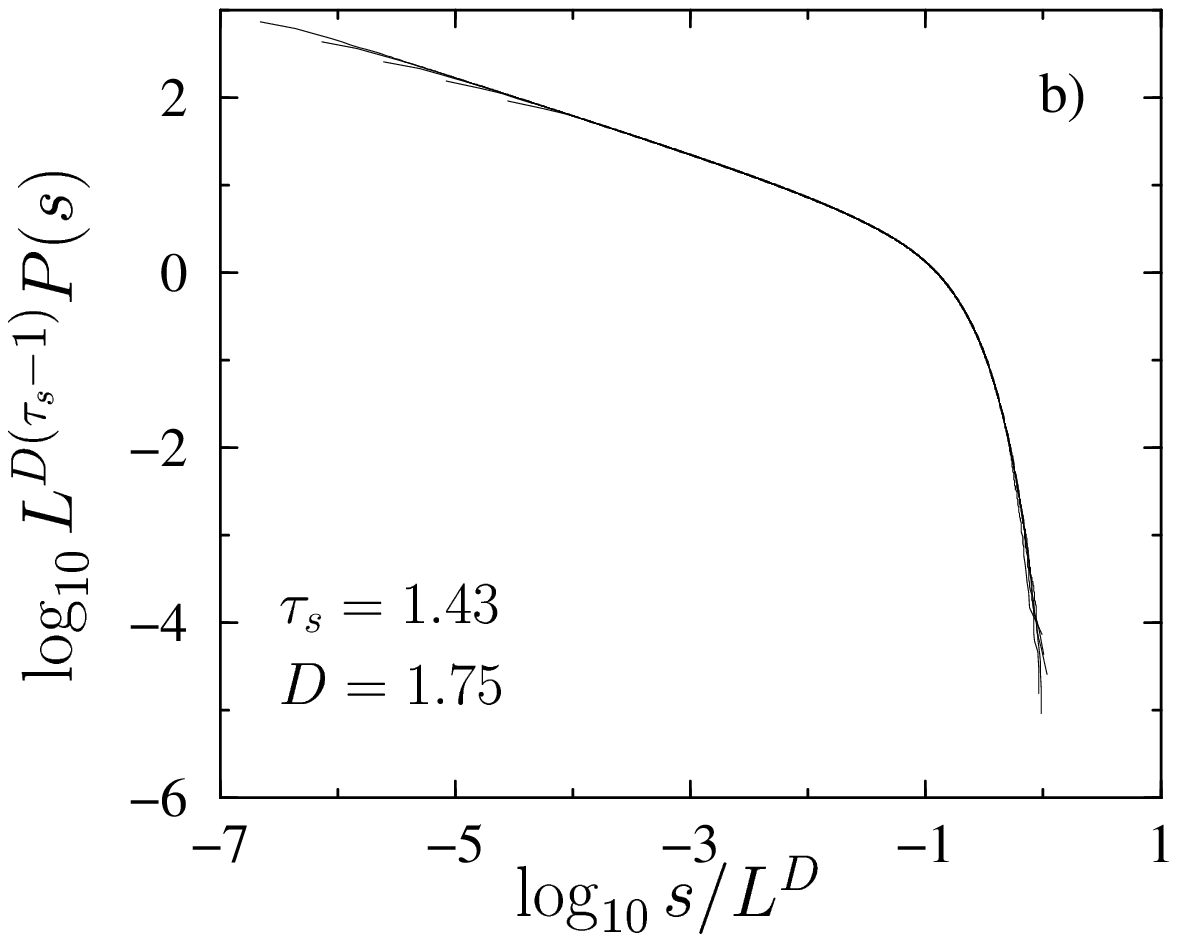, width=7cm}}
  \vspace*{0.25cm}
  \caption{Data collapse analysis of the integrated avalanche size
    distribution for the $d=2$ stochastic models with boundary
    dissipation a) ESDS and b) NESDS. System sizes are $L=400, 800,
    1600, 3200,$ and $6400$.}
  \label{fig:sizesL}
\end{figure}

\begin{figure}[t]
  \centerline{\epsfig{file=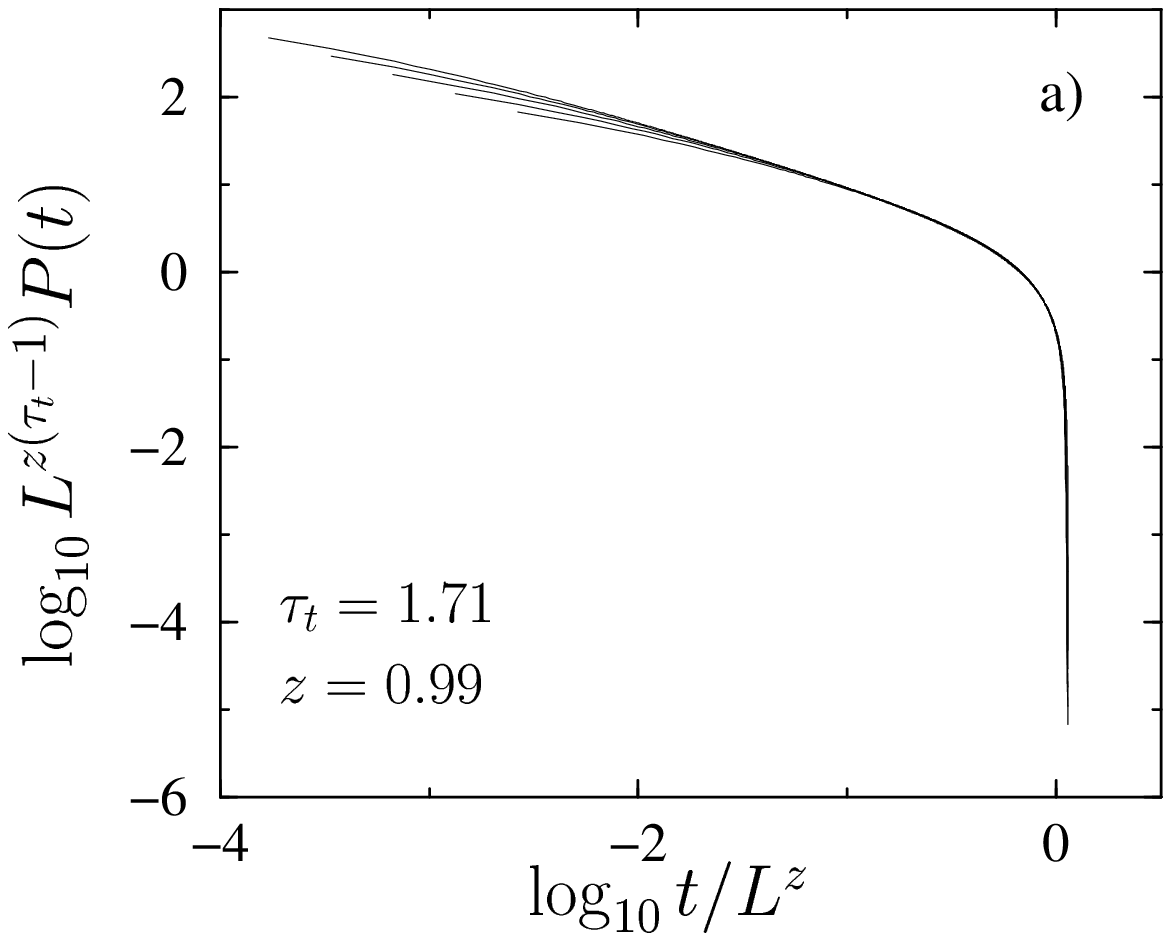, width=7cm} 
    \epsfig{file=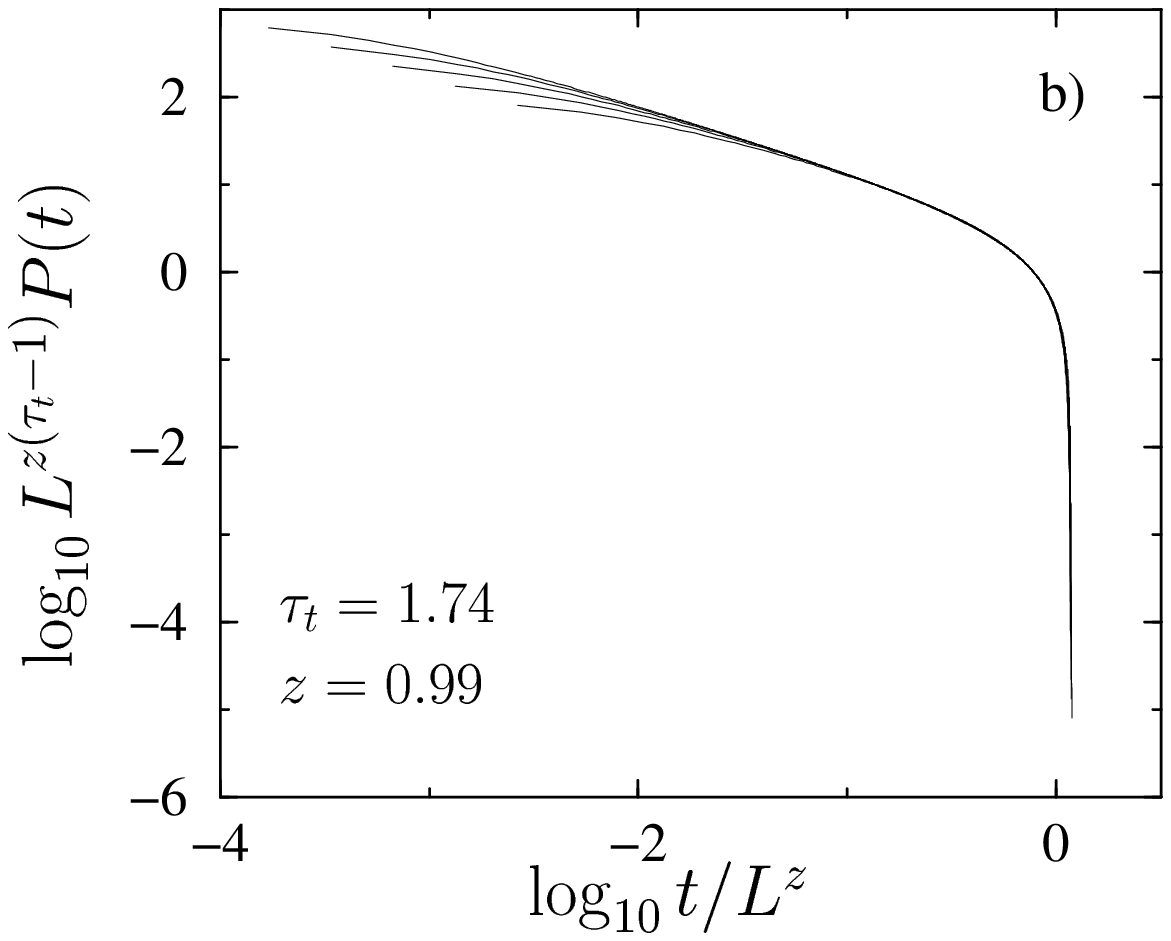, width=7cm}}
  \vspace*{0.25cm}
  \caption{Data collapse analysis of the integrated avalanche time
    distribution for the $d=2$ stochastic models with boundary
    dissipation a) ESDS and b) NESDS. System sizes are $L=400, 800,
    1600, 3200,$ and $6400$.}
  \label{fig:timesL}
\end{figure}

In Table~\ref{tableL} we report the exponents found for the DDS, ESDS,
and NESDS models in $d=2$.  Figure~\ref{fig:momentsL} shows the
moments $\sigma_s(q)$ and $\sigma_t(q)$. Figures~\ref{fig:sizesL} and
\ref{fig:timesL} plot the FSS data collapse for sizes and times,
respectively. The exponents obtained for the DDS are in perfect
agreement with the expected analytical results. This fact supports the
idea that the system sizes used in the present work allow to recover
the correct asymptotic behavior.  Results for the ESDS and NESDS are
identical within the error bars, pointing out that these two models
are in the same universality class. On the other hand, the obtained
exponents prove beyond any doubts that deterministic and stochastic
directed sandpile models do not belong to the same universality class.

We have also directly  computed the characteristic lengths in the 
parallel and transversal directions, $\xi_\parallel$ and $\xi_\perp$, 
as a function of the system size. The anisotropy of the system is 
reflected in the different definitions of both characteristic 
lengths. In this sense,
we define them with the same spirit as in directed percolation
\cite{kinzel83}.

Consider a given avalanche, labeled $\alpha$, that has started at the
site $(\xpa^{(0)}, \vxpe^{(0)})$, and has affected the set of {\em
  different} sites $\{(\xpa^{(i)}, \vxpe^{(i)})\}$, for $i=0\ldots
a-1$ (i.e., it has covered an area $a$). Let us define the quantities
\begin{equation}
  R_\parallel(\alpha)= \frac{1}{a}\sum_{i=1}^{a-1}
  |\xpa^{(0)}-\xpa^{(i)}| 
  \label{eq:corrdef1}
\end{equation}
and
\begin{equation}
  R_\perp^2(\alpha)= \frac{1}{a} \sum_{i=1}^{a-1}
  (\vxpe^{(0)}-\vxpe^{(i)})^2. 
  \label{eq:corrdef2}
\end{equation}
Furthermore, let us define $R_\parallel(a)$ and $R_\perp^2(a)$ as the
averages of the previous quantities, over all avalanches of the same
fixed area $a$. Let be $P(a)$ the probability of observing an
avalanche of area $a$. We define the correlation lengths by
\begin{equation}
  \xi_\parallel  = \frac{\sum_a R_\parallel(a) a P(a)}{\sum_a a P(a)},
  \qquad  
  \xi_\perp^2   = \frac{\sum_a R_\perp^2 (a) a P(a)}{\sum_a a P(a)}.
  \label{eq:corrdef}
\end{equation}
\begin{figure}[t]
  \centerline{\epsfig{file=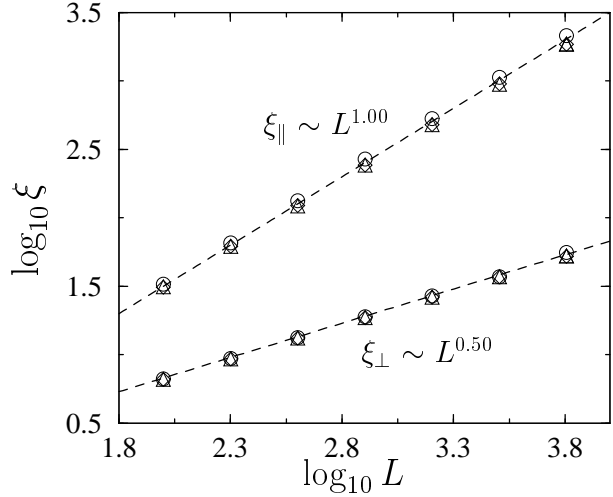, width=8cm}}
  \vspace*{0.5cm}
  \caption{Correlation lengths $\xi_\parallel$ and $\xi_\perp$ as a
    function of $L$ for the models with boundary dissipation DDS
    $(\circ)$, ESDS $(\triangle)$, and NESDS $(\diamond)$. The dashed
    lines are guides to the eye with slope $1.00$ and $0.50$.}
  \label{fig:correls}
\end{figure}
The different definitions \equ{eq:corrdef1} and \equ{eq:corrdef2} are
obviously due to the different nature of the avalanche spreading in
the directions $x_\parallel$ and $x_\perp$. In the former case, the
spreading is isotropic, and thus the second moment of the relative
distance distribution is needed to define a meaningful correlation
length. In the latter case, on the other hand, the spreading is always
in the direction of growing $x_\parallel$, and therefore the first
moment is sufficient.

The system being critical, both correlation lengths should scale with
the system size, defining the exponents $\nu_\parallel$ and
$\nu_\perp$ by
\begin{equation}
  \xi_\parallel  \sim L^{\nu_\parallel}, 
  \qquad  
  \xi_\perp \sim L^{\nu_\perp}.
\end{equation}
The affinity exponent, defined by 
\begin{equation}
  \xi_\perp \sim \xi_\parallel^{\zeta}
\end{equation}
is thus given by $\zeta=\nu_\perp / \nu_\parallel$.

We have calculated the correlations lengths in the models DDS, ESDS,
and NESDS, given by the definition \equ{eq:corrdef}. The results,
plotted in Fig.~\ref{fig:correls}, give the following dependence of
the correlation lengths with system size for all models:
\begin{equation}
  \xi_\parallel \sim L, \qquad  \xi_\perp \sim
  L^{1/2}.
\end{equation}
These relations define the exponents $\nu_\parallel=1$ and
$\nu_\perp=1/2$, and an affinity exponent $\zeta=1/2$.  It is
interesting to note that this exponent is independent of the
universality class of the model, defining a sort of super-universal
property of directed models.

As pointed out in Ref.\cite{pv99jpa}, the stochastic dynamics of SDS
models introduces multiple toppling events on the same site, which are
by definition absent in the deterministic case. This gives rise to
very a different avalanche structure, eventually reflected in the
different asymptotic critical behavior. It is worth remarking that the
universality class of SDS appears robust to modifications of the
stochastic microscopic dynamics as pointed out in Ref.\cite{vazq00},
where it is shown that modifications of SDS models with stochastic
toppling threshold still belong to the same universality class.
Recently, Paczuski and Bassler \cite{pacz00}, have proposed a
theoretical approach that allows the calculation of critical exponents
in directed models with multiple topplings. The analysis goes through
the mapping of the avalanche evolution into the dynamics of an
interface moving in a random medium, as also proposed in
\cite{pacz96,ala00}. This theoretical result gives the exponents
$\tau_s=10/7$ and $\tau_t=D=7/4$, in perfect agreement with the values
obtained by numerical simulations, Table~\ref{tableL}. The same
exponent values are also found in the approach of
Ref.~\cite{kloster00}.

\section{Numerical simulations with bulk dissipation}

In this Section we report results from computer simulations of
deterministic and stochastic sandpiles, performed with bulk
dissipation. In this case, dissipation is implemented as described in
Sec.~\ref{sec:models}. That is, in a system with periodic boundary
conditions, each toppling site has a probability $\epsilon/z_c$ of
losing an energy $z_c$, and a probability $1-\epsilon/z_c$ of
transferring it to its neighbors. The dissipation rates range from
$\epsilon=0.0016$ to $0.0512$, and the (fixed) system size considered
is $L=6400$. Statistical distribution functions have been computed
averaging over $10^7$ nonzero avalanches.

\begin{table}[b]
\begin{tabular}{lccccc}
  Model & $\tau_s$   & $\Delta_s$   & $\tau_t$   & $\Delta_t$ &
  $\zeta$ \\ 
\hline
DR & $4/3$ & $3/2$ & $3/2$ & $1$ & $1/2$ \\
DDS  &  $1.32(1)$ & $1.50(1)$ & $1.52(1)$ & $1.00(1)$ & $0.51(1)$\\
ESDS  &  $1.42(1)$ & $1.72(1)$ & $1.70(4)$ & $0.98(2)$ & $0.51(1)$ \\
NESDS  &  $1.43(1)$ & $1.75(1)$ & $1.70(5)$ & $0.99(2)$ & $0.50(1)$
\end{tabular}
\caption{Critical exponents for directed sandpiles with bulk
  dissipation in $d=2$. DR: Dhar and Ramaswamy's exact solution; DDS,
  deterministic directed model; ESDS and NESDS, stochastic directed
  models. Figures in parenthesis denote statistical uncertainties.} 
\label{tableeps}
\end{table}

\begin{figure}[t]
  \centerline{\epsfig{file=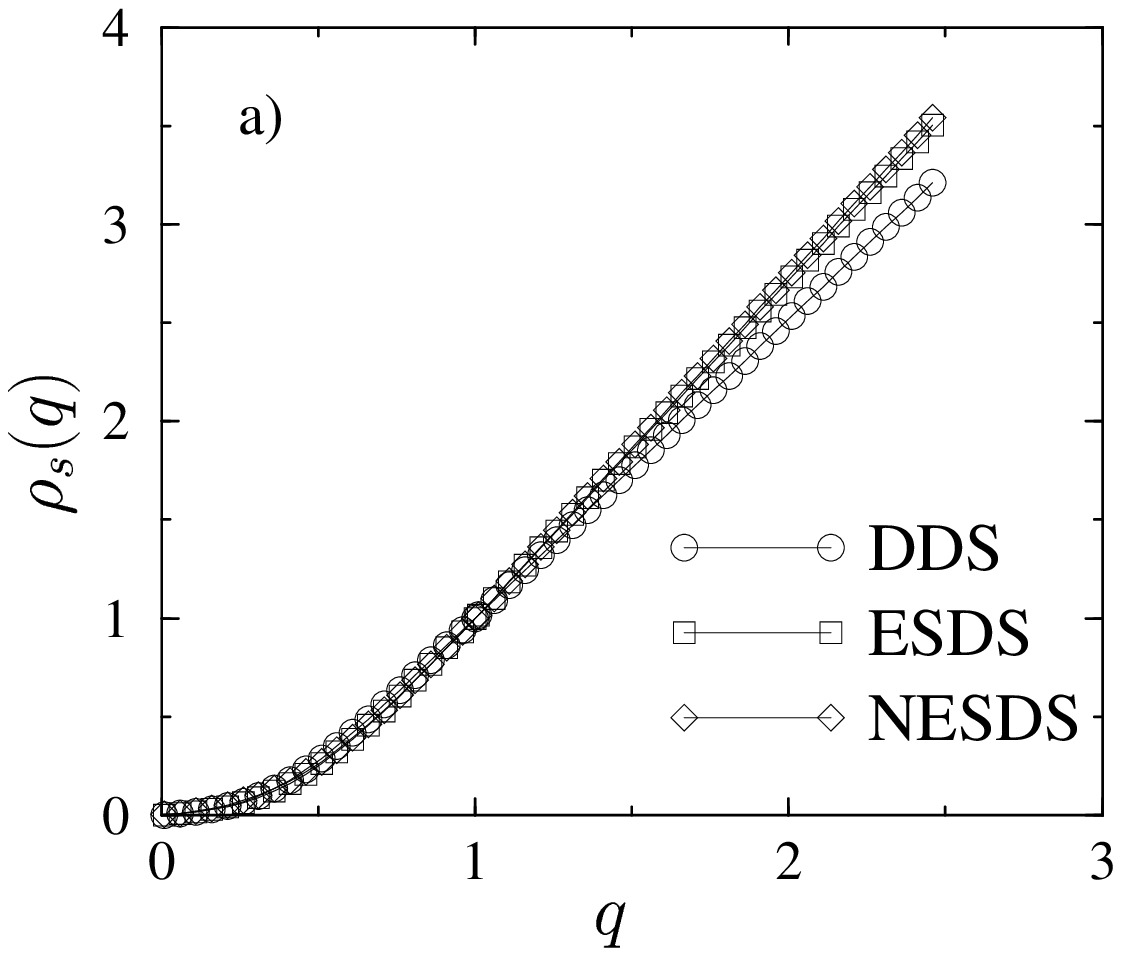, width=7cm} 
    \epsfig{file=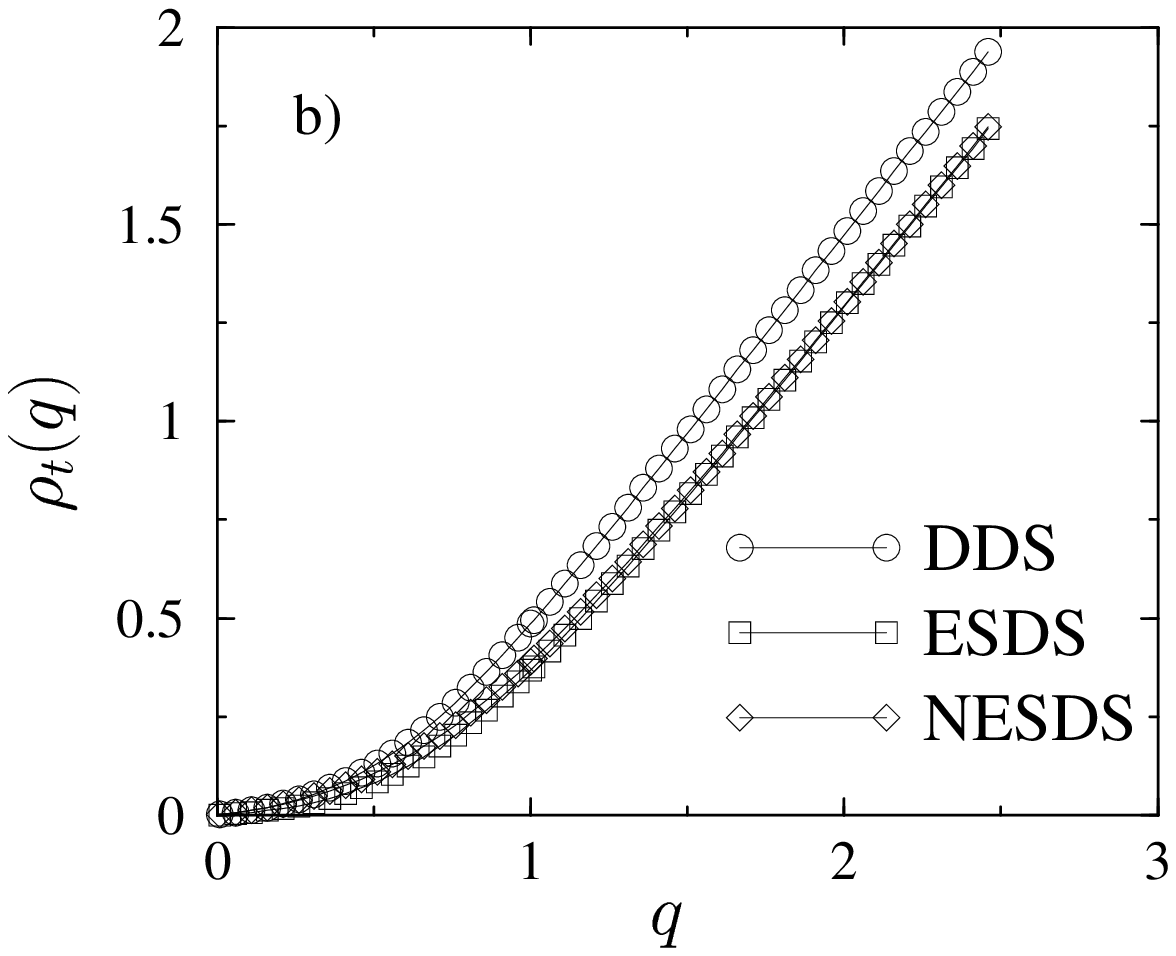, width=7cm}}
  \vspace*{0.25cm}
  \caption{Plot of (a) $\rho_s(q)$ and (b) $\rho_t(q)$ for the
    $d=2$ models DDS, ESDS, and NESDS with bulk dissipation.}
  \label{fig:momentseps}
\end{figure}

In the presence of bulk dissipation the characteristic sizes are
determined by the dissipation rate $\epsilon$, which defines the only
characteristic length in the system.  Approaching the limit
$\epsilon\to 0$, the avalanche characteristic size and time diverge as
$s_c\sim \epsilon^{-\Delta_s}$ and $t_c\sim \epsilon^{-\Delta_t}$,
respectively. It is also very easy to relate the mean avalanche size
to the dissipation rate $\epsilon$. On average, each added grain must
be dissipated in the evolution of the avalanche, resulting in
$\epsilon \left< s \right> =1$. This readily yields $\left< s \right>
= \epsilon^{-1}$.  In this case it is extremely important that the
characteristic length of the avalanche $\xi_\parallel$ is always
smaller than the size of the lattice used.  This allows us to study
only finite size effects introduced by the dissipation probability,
without spurious effects due to the finite lattice size.

The moment analysis can be straightforwardly generalized to systems
with bulk dissipation. In this case the role of the system size $L$ as
scaling parameter is played by the dissipation $\epsilon$. If the FSS
hypothesis holds, the $q$-th moment for, say the size distribution, has
an explicit dependence on the dissipation rate that reads:
\begin{equation}
 \left< s^q \right>_\epsilon \sim \epsilon^{-\Delta_s(q+1-\tau_s)} =
 \epsilon^{-\rho_s(q)}. 
\end{equation}  
The new moment $\rho_s(q)=\Delta_s(q+1-\tau_s)$ can be estimated by
linear regression in a log-log plot of $\left< s^q \right>_\epsilon$
as a function of $\epsilon^{-1}$. Once this moment is computed, the
exponent $\Delta_s$ is given by $\Delta_s=\partial \rho_s(q) /
\partial q$. The relation $\left< s \right> = \epsilon^{-1}$ imposes
$\rho_s(1)=1$, and from here, once known $\Delta_s$, we compute
$\tau_s$ using the relation $\rho_s(1)=\Delta_s(2-\tau_s)$.  Analogous
considerations allow to compute the exponents of the time
distribution, $\Delta_t$ and $\tau_t$. Finally, to check the exponents
with the data collapse technique, one must plot the rescaled functions
$P(s)\epsilon^{-\Delta_s \tau_s}$ as a function of
$s/\epsilon^{-\Delta_s}$ and $P(t)\epsilon^{-\Delta_t \tau_t}$ as a
function of $t/\epsilon^{-\Delta_t}$, respectively.

\begin{figure}[t]
  \centerline{\epsfig{file=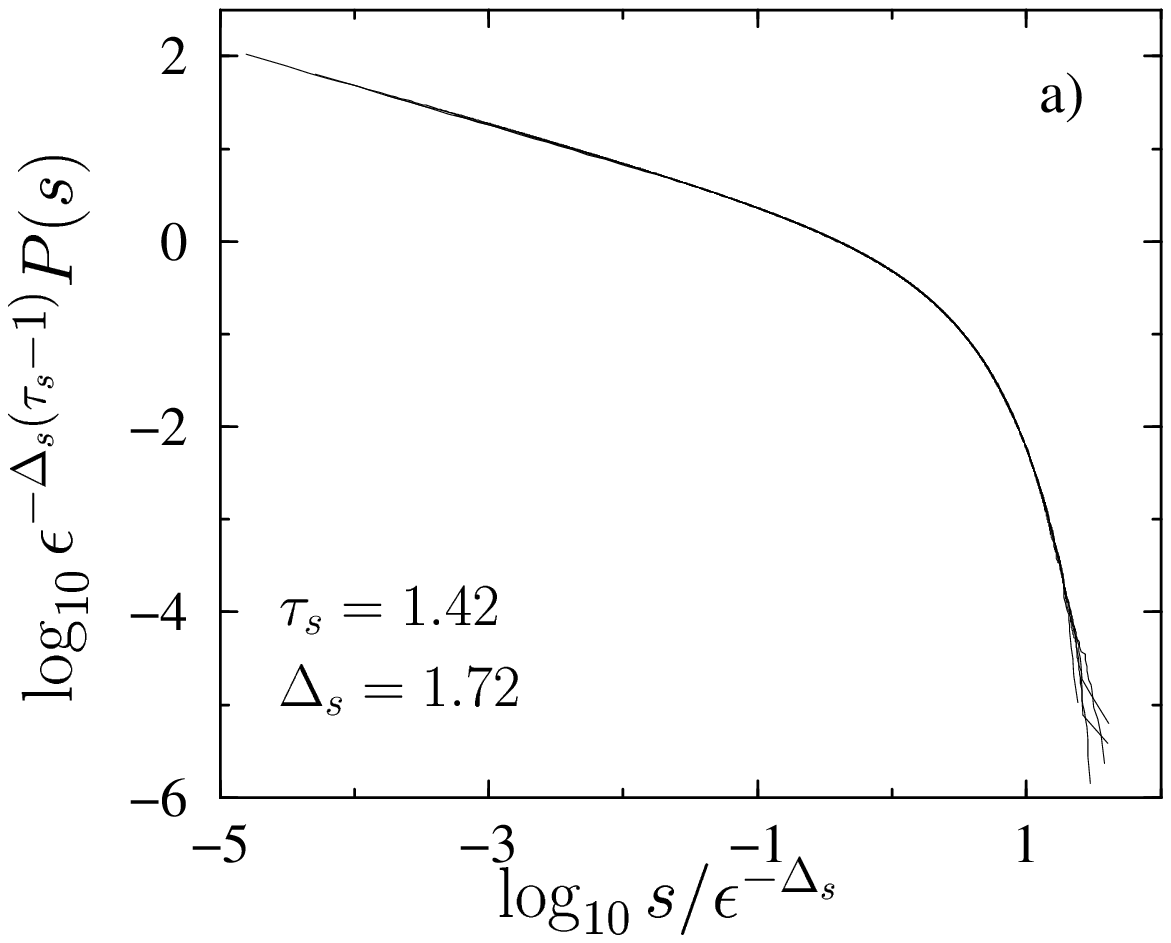, width=7cm} 
    \epsfig{file=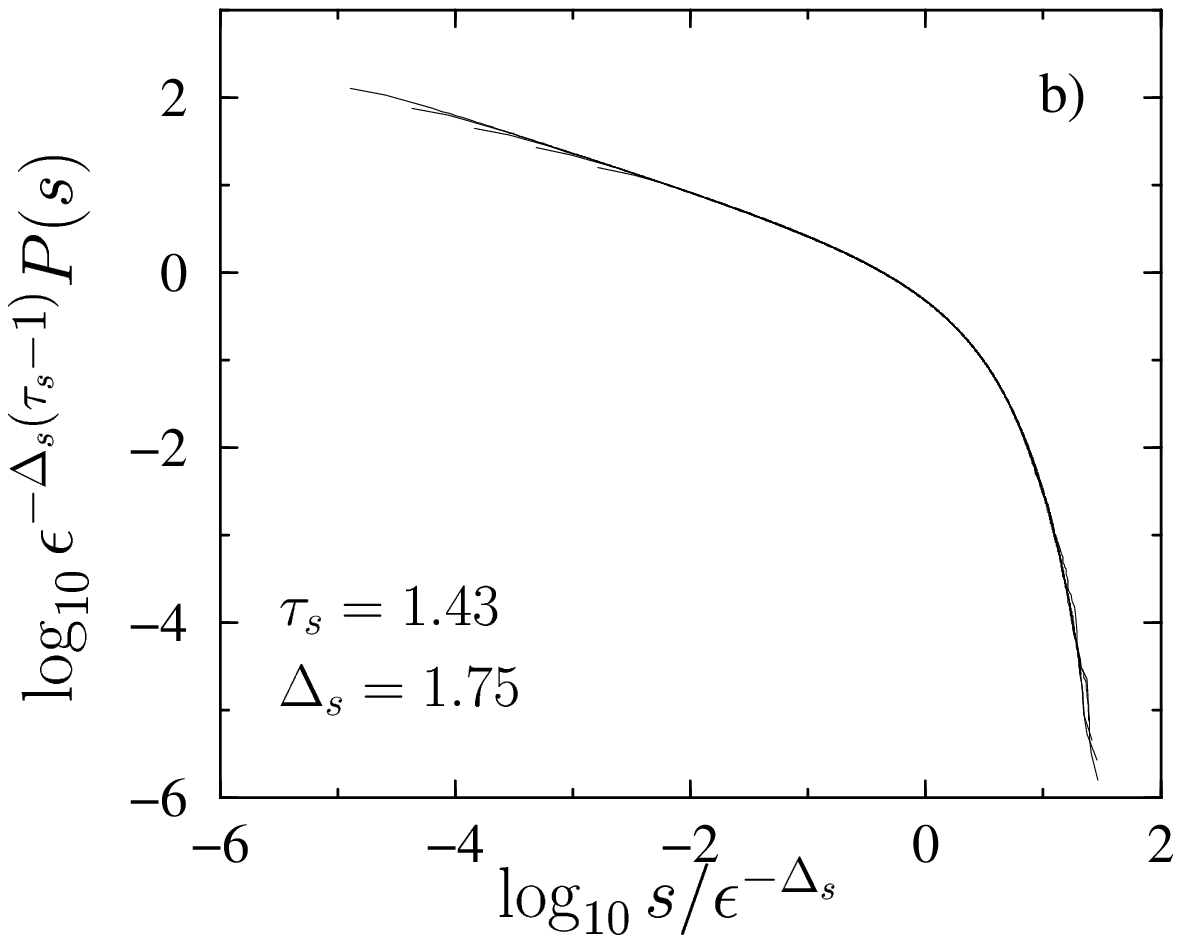, width=7cm}}
  \vspace*{0.25cm}
  \caption{Data collapse analysis of the integrated avalanche size
    distribution for the $d=2$ stochastic models with bulk
    dissipation a) ESDS and b) NESDS. Dissipations are
    $\epsilon=0.0256, 0.0128, 0.0064, 0.0032$, and $0.0016$.}
  \label{fig:sizeseps}
\end{figure}

\begin{figure}[t]
  \centerline{\epsfig{file=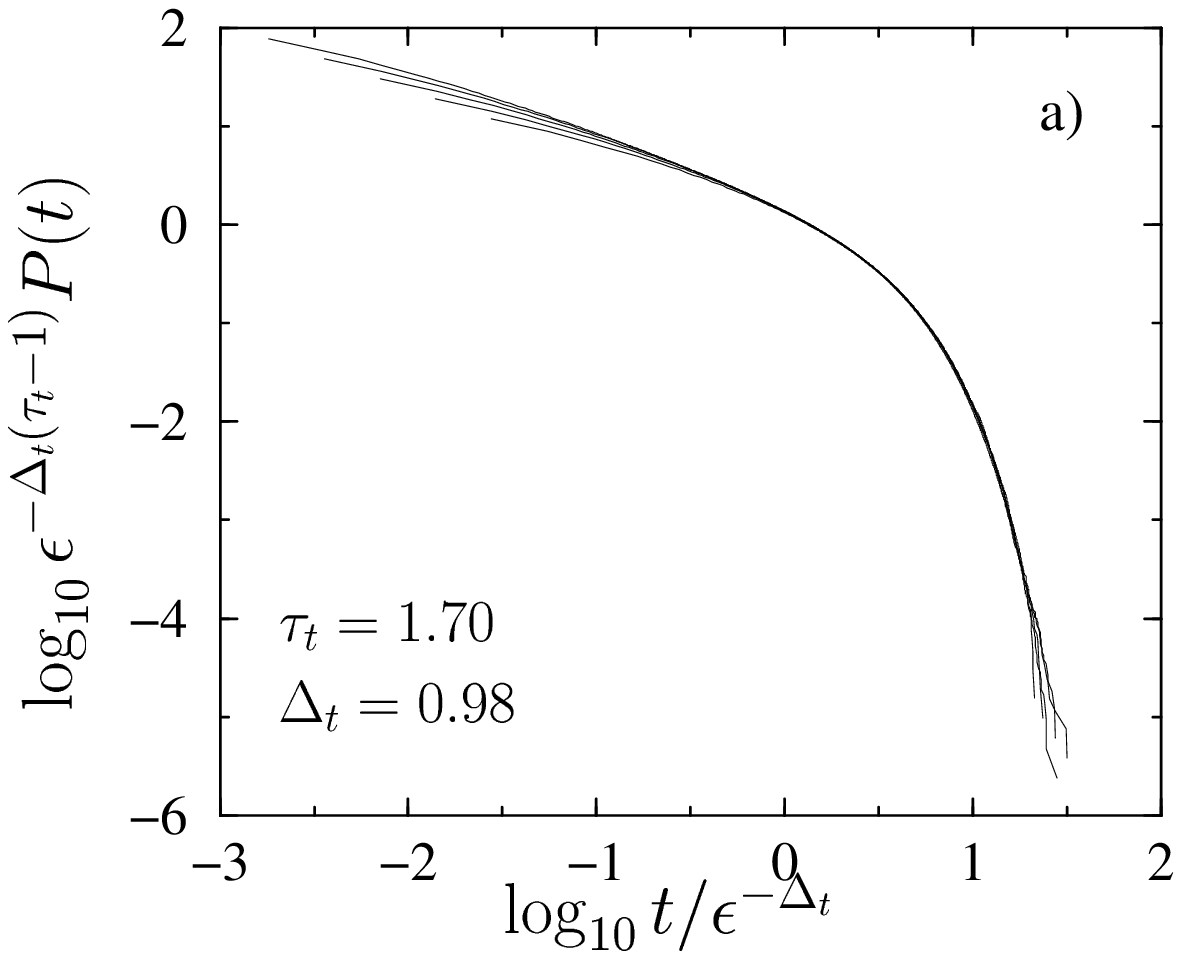, width=7cm} 
    \epsfig{file=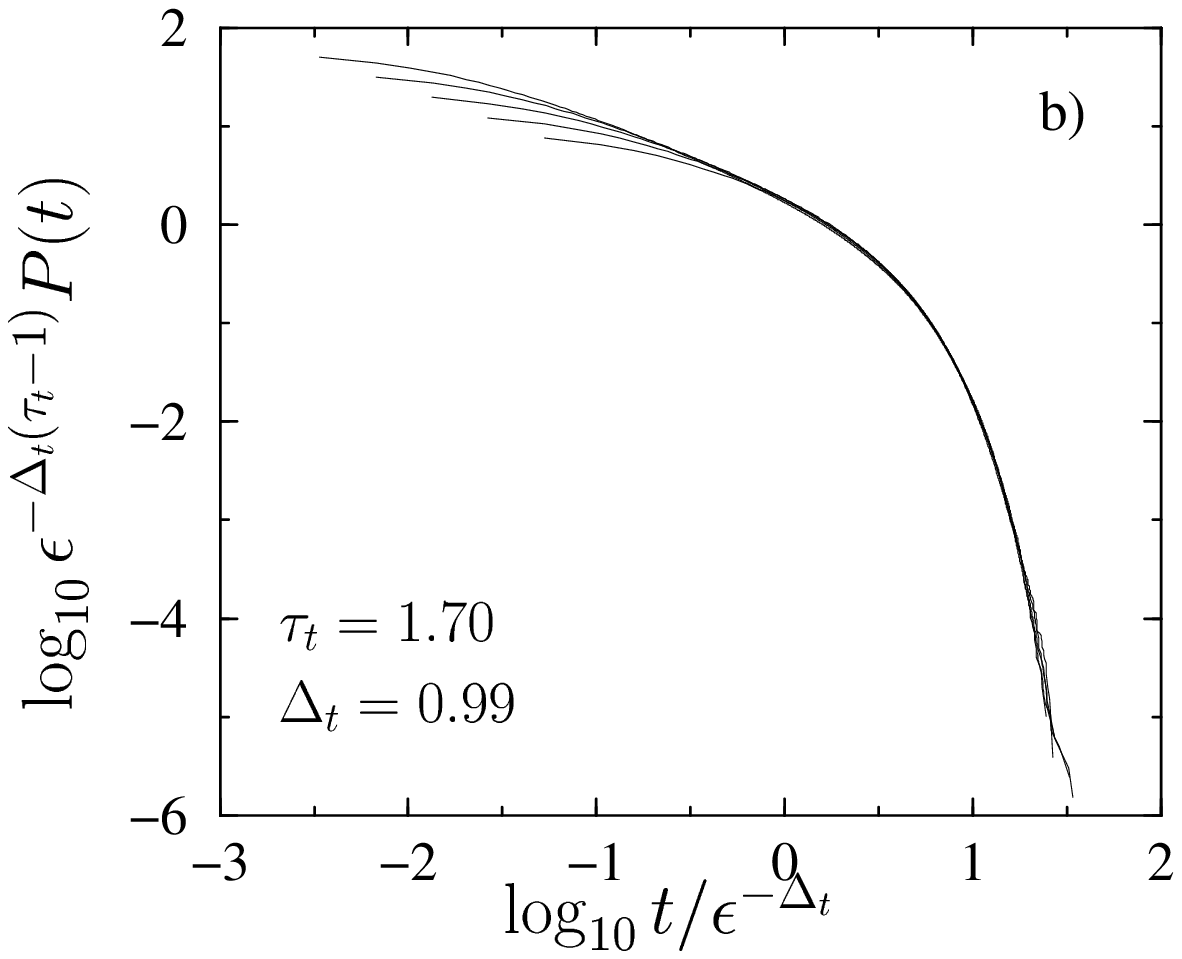, width=7cm}}
  \vspace*{0.25cm}
  \caption{Data collapse analysis of the integrated avalanche time
    distribution for the $d=2$ stochastic models with bulk
    dissipation a) ESDS and b) NESDS. Dissipations are
    $\epsilon=0.0256, 0.0128, 0.0064, 0.0032$, and $0.0016$.}
  \label{fig:timeseps}
\end{figure}

In Table~\ref{tableeps} we report the exponents computed in $d=2$ for
the directed models DDS, ESDS, and NESDS with bulk dissipation. The
corresponding moments $\rho_s(q)$ and $\rho_t(q)$ are shown in
Figures~\ref{fig:momentseps}, while Figs.~\ref{fig:sizeseps} and
\ref{fig:timeseps} plot the data collapse for sizes and times,
respectively.

\begin{figure}[t]
  \centerline{\epsfig{file=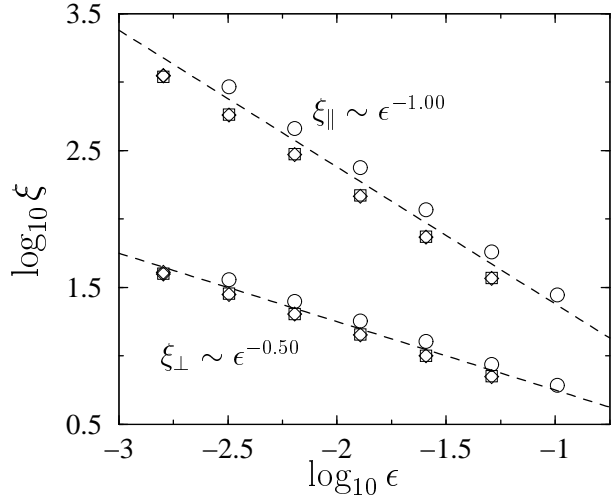, width=8cm}}
  \vspace*{0.5cm}
  \caption{Correlation lengths $\xi_\parallel$ and $\xi_\perp$ as a
    function of $\epsilon$ for the models with bulk dissipation DDS
    $(\circ)$, ESDS $(\triangle)$, and NESDS $(\diamond)$. The dashed
    lines are guides to the eye with slope $1.00$ and $0.50$.}
  \label{fig:correlseps}
\end{figure}

To conclude our analysis of directed sandpiles with bulk dissipation,
we have proceeded to compute the correlation length of the models.  In
this case, the scaling of the correlation lengths with vanishing
dissipation define the scaling exponents
\begin{equation}
  \xi_\parallel  \sim \epsilon^{-\nu_\parallel'}, 
  \qquad  
  \xi_\perp \sim \epsilon^{-\nu_\perp'}.
  \label{eq:correlexps}
\end{equation}
and an affinity exponent $\zeta=\nu_\perp' / \nu_\parallel'$.  Using
an analogous definition as in the case of boundary dissipation, we
compute the exponents $\nu_\parallel'=1$, $\nu_\perp'=1/2$, and
$\zeta=1/2$, as shown in Fig.~\ref{fig:correlseps}. That is, the
correlation length exponents are identical for both boundary and bulk
dissipation.  These results again imply an affinity exponent
$\zeta=1/2$ in all the models studied so far.

In view of these results, we have confirmation that the stochastic
models belong to a different universality class than the deterministic
directed sandpile.  These results also point out in a very clear way
that the critical behavior of models with boundary or bulk dissipation
is identical. In fact, all critical exponents $\tau_s, \tau_t, z,$ and
$\zeta$ are identical in both cases\cite{notascaling}.  This further
confirms the complete equivalence of both points of view with respect
to sandpiles and shows that, at least in the directed case, the open
boundary conditions usually implemented in simulations do not affect
the scaling behavior in a peculiar way. Of course, the open boundary
conditions breaks the translational invariance of the system, but in
the thermodynamic limit this effect is negligible for the asymptotic
critical behavior. Finally, these results validate theoretical
approaches in which it is assumed a homogeneous dissipation that is
much easier to treat analytically.

As a last observation it is worth remarking that also in this case, a
series of exponents such as $\zeta$ and $\nu_\perp'$ assume values
independently of the universality class of the model under study.
This sort of super-universality can be explained in terms of energy
conservation as we shall see in Sec. \ref{sec:conservation}.

\section{Numerical simulations of anisotropic models}

An important question to study in directed sandpile models is the
effect on the scaling properties of any amount of diffusion along the
preferred direction of transport $x_\parallel$.  One would expect that
the broken symmetry introduced by the preferential direction should
prevail on large scales, so that the dynamical scaling in directed and
simply anisotropic sandpiles become indistinguishable in the
thermodynamic limit. This fact hints towards the possibility of a
unique universality class for both directed and anisotropic sandpiles.
This universality class is determined uniquely by the lack of symmetry
along the $\xpa$ direction, and the presence or absence of stochastic
elements in the definition of the models.

\begin{figure}[t]
  \centerline{\epsfig{file=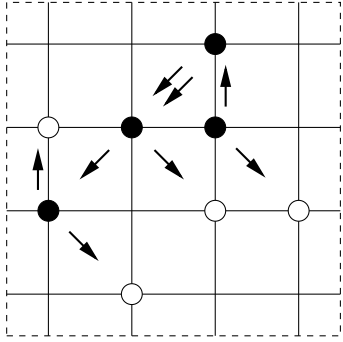, width=4.5cm}}
  \vspace*{0.5cm}
  \caption{Toppling rules in $d=2$ for an anisotropic sandpile. Filled
    circles represent active (toppling) sites; empty circles are
    stable sites. An active site sends one grain to two randomly
    chosen sites selected among the three downwards neighbors and the
    upward nearest neighbor.}
  \label{fig:rulesaniso}
\end{figure}

In order to test this conjecture, we have performed numerical
simulations of an anisotropic stochastic sandpile model, defined
according to the following rules: on a hyper-cubic lattice of size
$L$, we consider a model with threshold $z_c=2$. When a site topples,
it sends two grains of energy to two sites, randomly selected among
the $2d+1$ nearest and next-nearest neighbors on the hyper-plane
$\xpa+1$, {\em and} the nearest neighbor on the hyper-plane $\xpa-1$,
see Fig.~\ref{fig:rulesaniso}. The rules in this model are defined
non-exclusive, in such a way that the same site can receive the two
sand grain expelled by an active site.  The model is clearly
anisotropic, because the probability to transfer energy in the
downwards direction is three times larger that in the upwards
direction.  It would thus correspond to a non-exclusive stochastic
anisotropic sandpile (NESAS). We consider only the case of boundary
dissipation, performing simulations for sizes ranging from $L=100$ up
to $6400$, and averaging over $10^7$ nonzero avalanches.

\begin{figure}[t]
  \centerline{\epsfig{file=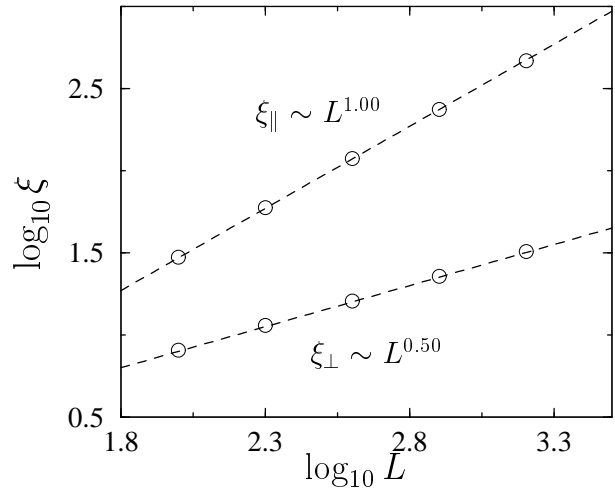, width=8cm}}
  \vspace*{0.5cm}
  \caption{Correlation lengths $\xi_\parallel$ and $\xi_\perp$ as a
    function of $L$ for the model with boundary dissipation NESAS. The
    dashed lines are guides to the eye with slope $1.00$ and $0.50$.}
  \label{fig:correlsaniso}
\end{figure}

In Fig.~\ref{fig:correlsaniso} we plot the correlation length
$\xi_\parallel$ and $\xi_\perp$, measured according to the rules given
in Eqs.~\equ{eq:corrdef}. We confirm the expectation that anisotropic
models have the same scaling properties, as regards the scaling of the
correlation lengths, as directed models with the same deterministic or
stochastic ingredients. We have also measured the exponents $\tau_s,
\tau_t, D$, and $z$ for this model, using the moment analysis
technique. The values found are $\tau_s=1.43(1)$, $D=1.75(1)$,
$\tau_t=1.72(2)$, $z=0.98(2)$. These results, compared with
Tables~\ref{tableL} and~\ref{tableeps}, show that this anisotropic
models belongs to the same universality class of the ESDS and NESDS
directed models, confirming the irrelevance of the diffusion along the
preferred direction $\xpa$.

\section{The role of conservation in sandpile models}
\label{sec:conservation}

We have seen in the preceding Sections that a subset of critical
exponents characterizing the critical behavior of directed and
anisotropic models have an interesting super-universal property; i.e.
they are independent of the universality class of the models.  In
order to understand this feature we perform a theoretical analysis
based on the conservation of energy, that is the basic symmetry in
standard sandpile automata. We shall see in the following that the
super-universal character of some critical exponents is dictated by
simple energy conservation considerations. The use of this approach
allows also to establish a relation between boundary and bulk
dissipation models by introducing an effective dissipation that depends
on the system size.

The avalanche dynamics in sandpile models is implicitly due to the
imposed infinite time scale separation between driving and dissipation
\cite{vz98,dvz98,tang88}. In order to devise a theory that can take
into account the symmetry introduced by the energy conservation, one
must first regularize the rules of the models in such a way that a
single time scale is ruling the dynamics. One way to do so is to
introduce a nonzero driving rate, defined as the probability per unit
time $h$ of a site to receive a grain of energy \cite{vz98,tang88}.
This driving rate plays the role of an external field and leads to the
SOC behavior in the limit $h\to0^+$.  On the other hand, given that
the toppling rules are conserved, energy can leave the system only at
the boundaries. Boundary dissipation is a natural choice in computer
simulations. However, it introduces undesirable complications due to
its singular character in a local theory. It is therefore convenient
to use an homogeneous effective dissipation $\epsilon$, defined as the
average energy lost in each toppling event.  As observed in previous
Sections, one can define models with periodic boundary conditions and
built-in bulk dissipation. When constructing the local theory for
models with open boundary conditions, the bulk dissipation $\epsilon$
amounts to an effective parameter that is to be related to the system
size $L$.

With all these ingredients, we are ready to formulate conservation of
energy as a continuous equation.  In sandpiles, we define the order
parameter $\rho_a$ as the density of active sites (i.e., whose height
$z \geq z_c$). The only dynamics in the model is obviously due to the
field $\rho_a(\vec{x},t)$, which is coupled to the local energy
density, $E(\vec{x},t)$ (i.e. the local density of sandgrains), which
enhances or suppresses the generation of new active sites.  A Langevin
description for sandpile automata is possible by considering the
dynamics of the local order-parameter field $\rho_a(\vec{x},t)$ in a
coarse-grained picture, bearing in mind that the energy density
$E(\vec{x},t)$ is a {\it conserved} field.  In
Refs.~\cite{dvz98,vdmz98}, in analogy with absorbing-state phase
transitions \cite{marro99,pacz}, a pair of coupled dynamical equations for
the fields $\rho_a(\vec{x},t)$ and $E(\vec{x},t)$ were proposed.  
In the following we elucidate the consequences of energy conservation and
we focus only on the latter equation. The interested reader can find
the full set of equations in Ref.~\cite{vdmz98}.  In the next
subsections we shall consider separately directed and anisotropic
models.

\subsection{Directed sandpiles}

We seek a continuous equation for the coarse-grained local density of
energy, $E(\vec{x},t)$. In the limit of zero driving and dissipation,
energy is conserved. Therefore, the evolution equation fulfilled by
the local field $E$ is:
\begin{equation}
  \frac{\partial E(\vec{x},t)}{\partial t} = - \vec{\nabla} \cdot
  \vec{J}_E - 
  \epsilon  \rho_a(\vec{x},t) + h(\vec{x},t) + 
  \eta_E(\vec{x},t).
\end{equation}
The first term simply represents the diffusion of energy; the second
term accounts for the dissipation that is associated with every
toppling event; the third term represents the external driving.
Finally, the last term is a source of stochastic noise, that accounts
for the randomness in the flow of energy. The noise term can be
generated by the toppling rules in a stochastic model, or by the
initial conditions plus the random driving in a deterministic model.
We will require the noise to have zero average:
\begin{equation}
  \left< \eta_E(\vec{x},t) \right> = 0.
\end{equation}
The noise correlator $\left< \eta_E(\vec{x},t) \eta_E(\vec{x}',t'
\right>$ is of fundamental importance for the determination of
universality classes and the critical behavior of the order parameter.
However, for our present purposes we do not need precise knowledge of
its analytical form (for a detailed discussion, see
Refs.~\cite{dvz98,vdmz98}).

The current can be constructed by appealing to the symmetries of the
model. The transport of energy is due to topplings. These are isotropic
along the transversal direction $\vxpe$, therefore the current along
this direction will be proportional to the {\em gradient} of the
density of active sites. In the preferred direction, on the other
hand, all the energy is transferred downwards; therefore, the current
in this direction must be proportional to the {\em density} of active
sites.  The final form of the current is then
\begin{equation}
  \vec{J}_E(\vec{x},t) = - D_\perp \vec{\nabla}_\perp
  \rho_a(\vec{x},t) + 2 \lambda   \rho_a(\vec{x},t) \vec{e}_\parallel.  
\end{equation}
Plugging this expression into the equation for the energy, we have the 
final result:
\begin{equation}
  \frac{\partial E(\vec{x},t)}{\partial t} =  D_\perp \nabla^2_\perp
  \rho_a(\vec{x},t) - 2 \lambda \partial_\parallel \rho_a(\vec{x},t) -
  \epsilon  \rho_a(\vec{x},t) + h(\vec{x},t) +   \eta_E(\vec{x},t),
  \label{eq:finaldirected}
\end{equation}
where the symbol $\partial_\parallel$ stands for the partial
derivative $\partial / \partial \xpa$.  This is the general
conservation equation for any directed sandpile model. It is worth
remarking at this point that the energy field is a static field, in
the sense that energy diffuses only if active sites are present in the
system. This is intuitively understood in sandpile models, where
energy (sand) grains diffuse only from toppling sites.

To analyze the consequences of Eq.~\equ{eq:finaldirected}, it proves
useful to define the susceptibility $\chi(\vec{x},t)$ \cite{vdmz98}:
\begin{equation}
  \chi(\vec{x}-\vec{x}',t-t') = \left< \frac{\delta
  \rho_a(\vec{x},t)}{\delta h(\vec{x}',t')} \right>_\eta, 
  \label{eq:suscept}
\end{equation}
where the symbol $\left< \right>_\eta$ denotes an average over the
noise distribution. By definition, the susceptibility measures the
average increase in the number of active sites due to an impulsive
perturbation, that is, to the addition of a single energy grain. Since
we measure the size of the avalanches by the total number of
topplings, the average avalanche size is given by
\begin{equation}
 \left< s \right> = \int d^d x \, dt \, \chi(\vec{x},t).
\end{equation}

Taking the functional derivative of Eq.~\equ{eq:finaldirected} and
averaging over time and noise, we obtain, in the limit $t\to\infty$,
in which the sandpile is in a stationary state with constant average
energy, the following equation for the static susceptibility:
\begin{equation}
  D_\perp \nabla^2_\perp \chi(\vec{x}) - 2 \lambda  \partial_\parallel 
  \chi(\vec{x}) - \epsilon \chi(\vec{x}) = -\delta^{(d)} (\vec{x}). 
  \label{eq:suscepteq}
\end{equation}
This equation can be easily solved in Fourier space. Defining the
transformation 
\begin{equation}
  \chi(\xpa, \vxpe) = \frac{1}{(2\pi)^d} \int d^{d-1} k \, d q
  \, \chi( q, \vec{k}) e^{i\vec{k}\cdot\vxpe}  e^{iq\xpa} 
\end{equation}
and substituting into Eq.~\equ{eq:suscepteq}, we obtain the solution
\begin{equation}
  \chi(q, \vec{k}) = \frac{1}{D_\perp k^2 +2i\lambda q +\epsilon},
  \label{eq:susceptq}
\end{equation}
which yields the susceptibility in real space
\begin{equation}
  \chi(\xpa, \vxpe) = \frac{1}{(2\pi)^d} \int d^{d-1} k
  e^{i\vec{k}\cdot\vxpe} \int_{-\infty}^{\infty} d q
  \frac{e^{iq\xpa}}{D_\perp k^2 +2i\lambda q +\epsilon}.  
\end{equation}
The integral in $q$ is easily performed by residues. The remaining
$d-1$ integrals in $\vec{k}$ become then decoupled Gaussian integrals
that yield the result, setting $D_\perp=1$:
\begin{equation}
  \chi(\xpa, \vxpe) = \frac{1}{2 \lambda} 
  \left( \frac{\lambda}{2 \pi}\right)^{(d-1)/2}  
  \xpa^{(1-d)/2}   e^{-\xpa \epsilon / 2 \lambda} e^{-\lambda
  \xpe^2 /2 \xpa}.  
  \label{eq:directedso}
\end{equation}
Eq.~\equ{eq:directedso} can be conveniently rewritten into the scaling
form
\begin{equation}
  \chi(\xpa, \vxpe) = \xpa^{(1-d)/2}
  \Gamma\left(\frac{\xpa}{\xi_\parallel} , \frac{\xpe}{\xi_\perp} \right),
\end{equation}
where $\Gamma$ is a cut-off function that decreases exponentially in
both its arguments.  Comparing this last expression with
Eq.~\equ{eq:directedso}, we can identify the parallel and transversal
correlation lengths:
\begin{equation}
  \xi_\parallel \sim \epsilon^{-1}, \qquad  \xi_\perp \sim
  \epsilon^{-1/2}.
  \label{eq:correls}
\end{equation}
In more general terms, if we define the exponents $\nu_\parallel'$ and
$\nu_\perp'$ by Eqs.~\equ{eq:correlexps}, then we have for directed
sandpiles $\nu_\parallel'=1$ and $\nu_\perp'=1/2$. From these last
expressions, we can read off a first exact result for directed
sandpiles: the avalanches produced in those models are elongated, with
characteristic length in the parallel and transversal directions
related by an affinity exponent $\zeta=1/2$.  It is very important to
stress that these results are independent of the particular model
consider and of the dimensionality $d$ of the system, dictated only by
the energy balance in the stationary state.

We can use the result \equ{eq:correls} to relate the effective bulk
dissipation with the system size in a model with open boundary
conditions. To sustain a steady state with constant average energy,
avalanches must reach the bottom boundary in order to be able to
dissipate. This means that the characteristic length of the avalanches
in the parallel direction must be proportional to the system size
$\xi_\parallel \sim L$. We have therefore that in boundary dissipation
models we can define an effective dissipation rate $\epsilon$ that is
related with the system size by:
\begin{equation}
  \epsilon \sim L^{-1}.
  \label{eq:correspondence}
\end{equation}
From this relations we easily find that $\Delta_s=D$ and $\Delta_t=z$.
These identities are recovered in numerical simulations (see
Tables~\ref{tableL} and~\ref{tableeps}).  Finally, from
Eq.~\equ{eq:susceptq}, we can recover the well-known result linking
the system size and the average avalanche size, $\left< s \right> =
\chi(q=0, \vec{k}=0) \sim \epsilon^{-1} \sim L$
\cite{kad89,dhar89,tsuchiya99}.  That is, for any directed model, the
average avalanche size must scale as the inverse of the dissipation,
in the case of model with bulk dissipation, or as the size of the
system along the preferred direction of transport for sandpiles
defined with boundary dissipation.

\subsection{Anisotropic sandpiles}
\label{sec:anisotheory}

Having completed the analysis of directed sandpiles, we turn our
attention to the more complex case of anisotropic sandpiles. In this
kind of model, the transport of energy is not strictly directed in the
parallel direction, but is simply stronger in the direction
$+x_\parallel$ than in the opposite direction $-x_\parallel$. The
presence of backwards flow allows the possibility of diffusion in the
preferred direction, and thus the equation for the conservation of
energy becomes in this case
\begin{equation}
  \frac{\partial E(\vec{x},t)}{\partial t} =  D_\perp \nabla^2_\perp
  \rho_a(\vec{x},t) + D_\parallel \partial_\parallel^2
  \rho_a(\vec{x},t) - 2 \lambda  \partial_\parallel \rho_a(\vec{x},t) -
  \epsilon  \rho_a(\vec{x},t) + h(\vec{x},t) +   \eta_E(\vec{x},t). 
  \label{eq:finalaniso}
\end{equation}
Even though it is straightforward to obtain Eq.~\equ{eq:finalaniso} by
symmetry arguments, it is instructive to derive it also starting from
the anisotropy of the microscopic rules. To this end, let us consider
a one dimensional model in which active sites, when toppling, send a
fraction of energy $p$ to the left neighbor, and a fraction $(1-p)$ to
the right neighbor, with $0 \leq p \leq 1$. Let us
coarse-grain the line into cells of a certain (small) size. The
variation of energy of the $i$-th cell after a parallel updating will
be given by
\begin{equation}
  \Delta E^{(i)} \simeq -\rho_a^{(i)} + (1-\epsilon)p\rho_a^{(i+1)} +
  (1-\epsilon)(1-p)\rho_a^{(i-1)}, 
\end{equation}
where $\rho_a^{(i)}$ is the density of active sites in the $i$-th
cell. The proportionality stems from the fact that the actual
contribution comes only from the boundary sites in the cell.  This
last equation can be rewritten
\begin{equation}
  \Delta E^{(i)} \simeq -\epsilon\rho_a^{(i)} - (1-\epsilon) p
  \left[\rho_a^{(i)}-\rho_a^{(i+1)}\right]    -   (1-\epsilon) (1-p)
  \left[\rho_a^{(i)}- \rho_a^{(i-1)}\right]. 
\end{equation}
For $p=0$ or $p=1$ (for a strictly directed model), only one term
remains and, after passing to a continuous formulation, we recover
Eq.~\equ{eq:finaldirected}, restricted to the preferred direction.
For $p\neq0, 1$, and after performing some algebraic manipulations, we
obtain
\begin{eqnarray*}
  \Delta E^{(i)} &\simeq& -\epsilon\rho_a^{(i)} + (1-\epsilon)
    \left[\rho_a^{(i+1)} + \rho_a^{(i-1)} - 2 
    \rho_a^{(i)} \right] \\
  &-&  (1-\epsilon) p \left[\rho_a^{(i)}- \rho_a^{(i-1)}\right]+
  (1-\epsilon) (1-p) \left[\rho_a^{(i)}- \rho_a^{(i+1)}\right] \\
  &\simeq&   - \epsilon\rho_a^{(i)} +(1-\epsilon)p
    \partial_x^2\rho_a^{(i)} + (1-\epsilon)p \partial_x \rho_a^{(i)} +  
   (1-\epsilon)(1-p) (-\partial_x \rho_a^{(i)})\\
   &=& - \epsilon\rho_a^{(i)} +(1-\epsilon)p \partial_x^2\rho_a^{(i)} -
    (1-\epsilon)(1-2p) \partial_x \rho_a^{(i)}, 
\end{eqnarray*}
where in the third line we have passed to the continuum limit,
defining the discretized first derivative
$\partial_x\rho_a^{(i)}=\rho_a^{(i)} - \rho_a^{(i+1)}$ and second
derivative $\partial_x^2\rho_a^{(i)}=\rho_a^{(i+1)} + \rho_a^{(i-1)} -
2 \rho_a^{(i)}$. For $p=1/2$ we recover, as expected, an isotropic
equation. For all other values of $p$, the equation for the energy
will be effectively given by \equ{eq:finalaniso}, with the
phenomenological parameter $\lambda$ given at the microscopic level by
the expression $(1-\epsilon)(1/2-p)$.

From Eq.~\equ{eq:finalaniso}, we can obtain the corresponding equation 
for the susceptibility. The solution in Fourier  space is readily
found to be
\begin{equation}
  \chi(q,\vec{k}) = \frac{1}{D_\perp k^2 + D_\parallel q^2 + 2i\lambda
  q + \epsilon}. 
\end{equation}
Upon integration over $\vec{k}$ and $q$, one obtains the expression in
real space
\begin{equation}
  \chi(\xpa, \vxpe) = \frac{1}{(2\pi)^d} \int d^{d-1} k \,
  e^{i\vec{k}\cdot\vxpe}  \int_{-\infty}^{\infty} d q  \,
  \frac{e^{iq\xpa}}{D_\perp k^2 +D_\parallel q^2 +  2 i\lambda q 
    +\epsilon}.  
  \label{eq:bigeq}
\end{equation}
This last integral can be performed analytically in $d=1$ and $2$ (see
Appendix). For $d>2$, even though we do not have a closed expression,
we can obtain the leading scaling behavior.  To simplify the
calculations, we set, without lack of generality,
$D_\perp=D_\parallel=1$. The integration in $q$ is done by the method
of the residues (see Appendix). The integration of the $\vec{k}$
angular part \cite{formulitas} yields
\begin{equation}
  \chi(\xpa, \vxpe) = \frac{1}{2}
  \left(\frac{\gamma}{2\pi}\right)^{\nu+1} \xpe^{-\nu} 
  \int_0^\infty dz \, z^{\nu +1} J_\nu(\gamma \xpe z) \,
  \frac{e^{-\xpa(\gamma\sqrt{1+z^2} -\lambda)}}{(1+z^2)^{1/2}}.
  \label{eq:chiexact}
\end{equation}
Here, $J_\nu(z)$ is the first kind Bessel function of order $\nu$, and
we have defined the constants $\nu=(d-3)/2$ and
$\gamma=(\lambda^2+\epsilon)^{1/2}$.  We are interested in the
behavior of this integral for large distances, that is, in the limit
$\xpa\gg\xpe\gg1$. In this limit, the weight of the integral is given
by the region of small $z$, since the exponential suppresses large
values.  We can then approximate the integral in the interval $0<z<1$
and perform a Taylor expansion of the square root in the exponential
and the denominator. In the denominator, we readily have
$(1+z^2)^{1/2}\simeq1$. The term in the exponential, however, contains
a constant term, and must be therefore expanded up to second order:
\begin{equation}
  -\xpa(\gamma\sqrt{1+z^2}-\lambda) \simeq -\xpa(\gamma[1+z^2/2]
  -\lambda) = -\xpa(\gamma-\lambda) -\xpa \gamma z^2/2.
\end{equation}
In the  limit $\epsilon\to0$,  we have $\gamma\simeq\lambda$,  and the
constant $\gamma-\lambda$ can be expanded to give:
\begin{equation}
  \gamma-\lambda = (\lambda^2+\epsilon)^{1/2} - \lambda \simeq
  \lambda\left(1+\frac{\epsilon}{2\lambda^2} \right) - \lambda = 
  \frac{\epsilon}{2\lambda}.
\end{equation}
Substituting these approximations into Eq.~\equ{eq:chiexact}, we are
led to the expression
\begin{equation}
  \chi(\xpa, \vxpe) \simeq  \frac{1}{2\lambda}
  \frac{1}{(2\pi)^{\nu+1}} \xpe^{(1-d)} e^{-\xpa \epsilon / 2 \lambda} 
  \int_0^\infty dy \ y^{\nu+1} J_\nu(y) \, e^{-\frac{\xpa}{2 \lambda 
  \xpe^2} y^2},
  \label{eq:duable}
\end{equation}
where we have performed the change of variables $y=\gamma\xpe z$ and
extended again the upper limit of the integral to infinity (which is
allowed given its exponential convergence). The integral in
Eq.~\equ{eq:duable} yields \cite{formulitas}:
\begin{equation}
  \chi(\xpa, \vxpe) \simeq  \frac{1}{2 \lambda} \left(\frac{\lambda}{2
      \pi}\right)^{\nu+1} \xpa^{(1-d)/2} e^{-\xpa \epsilon/ 2 \lambda} 
  e^{- \lambda \xpe^2 / 2 \xpa},
\end{equation}
which as usual we can write in the scaling form,
\begin{equation}
  \chi(\xpa, \vxpe) = \xpa^{(1-d)/2}
  \Gamma\left(\frac{\xpa}{\xi_\parallel} , \frac{\xpe}{\xi_\perp}
  \right).
\end{equation}
From here, the correlation exponents read $\nu_\parallel'=1$ and
$\nu_\perp'=1/2$, as in the directed case. This implies again an
affinity exponent $\zeta=1/2$.

The conclusion of the lengthy calculations developed in this section
is that the presence of any amount of diffusion along the preferred
direction of a directed sandpile model is completely irrelevant. As
soon as there is anisotropy in a model (in our mathematical
formulation, when $\lambda \neq 0$, however small), it takes over and
places the model in the universality class of completely directed
sandpiles.  In particular, we recover the result $\left< s \right>
\sim L$ for any anisotropic sandpile, in agreement with the numerical
results in Ref.~\cite{kad89} and the analytic results of
Ref.~\cite{tsuchiya99}. We remark, however, that our results do not
rely in a particular model like those of \cite{tsuchiya99}, but only
on symmetry arguments, and are therefore of a broader generality.

\section{Conclusions}

\begin{figure}[t]
  \centerline{\epsfig{file=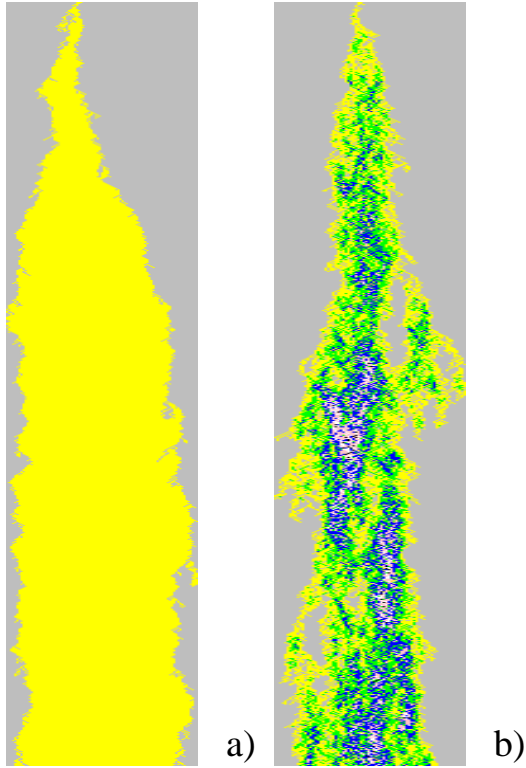, width=7cm}} 
  \vspace*{0.25cm}
  \caption{Color plots of the local density of topplings in two
  avalanches of size 50 000 for the  (a) DDS and (b) ESDS
  models. White stands for a sigle toppling per site; black represents
  the maximum number of topplings.}
  \label{fig:densityplot}
\end{figure}

In this paper we have presented a detailed numerical analysis of
deterministic and stochastic directed sandpile models. We find
definitive evidence for the existence of a new universality class,
embracing directed sandpile models with stochastic rules. The origin
of the different critical behavior can be traced back to the presence
of multiple topplings in the latter case. An example of this feature
is provided in Fig.~\ref{fig:densityplot}, where we plot the local
density of topplings in two avalanches corresponding to the DDS and
ESDS models. From this figure it becomes evident that the stochastic
dynamics induces multiple toppling events, which are forbidden in the
deterministic models. This feature has been fruitfully exploited in 
Ref.~\cite{pacz00} to obtain an analytical solution of the stochastic
model. 

We have also studied the case of directed sandpiles with bulk
dissipation. In this case, our results prove that the critical
behavior is unchanged. This points out that the boundary dissipation
does not play any particular role in the development of the critical
behavior in directed sandpiles. 

Finally, numerical results indicate that some critical exponents show
a super-universal nature, assuming the same values independently of
the universality class. We provide an analytical explanation of this
feature by means of a continuous phenomenological equation that takes
into account the energy balance condition imposed by the dynamical
rules in sandpile models.

\section*{Acknowledgements}

This work has been supported by the European Network under Contract
No.  ERBFMRXCT980183. We thank D. Dhar, R. Dickman, M. A. Mu{\~n}oz, A.
Stella, and S. Zapperi for helpful comments and discussions.

\appendix
\section*{}
In this Appendix, we work out the exact value of the integral
\equ{eq:bigeq}, giving the static susceptibility for anisotropic
sandpiles, in $d=1$ and $2$. We take again, without lack of
generality, $D_\perp=D_\parallel=1$.

\subsection*{Static susceptibility in $d=1$}
\label{residues}

In $d=1$, Eq.~\equ{eq:bigeq} reduces to the simple form
\begin{equation}
  \chi(\xpa) = \frac{1}{2\pi} \int_{-\infty}^{\infty} d q  \,
  \frac{e^{iq\xpa}}{q^2 +  2 i\lambda q +\epsilon}.  
\end{equation}
This integral can be done by the method of the residues. The zeroes of
the denominator are
\begin{equation}
  q_\pm = i [\pm (\lambda^2 + \epsilon)^{1/2} - \lambda].
\end{equation}
Closing the contour of the integral arround the upper complex plane,
we obtain
\begin{equation}
  \chi(\xpa) = \frac{1}{2} \, \frac{e^{- \xpa (\sqrt{\lambda^2 + \epsilon} -
  \lambda)}}{\sqrt{\lambda^2 + \epsilon}}.
\end{equation}
We are interested in the limit $\epsilon\to0$. The denominator gives
$\sqrt{\lambda^2 + \epsilon} \simeq \lambda$, while the argument in
the exponential yields
\begin{equation}
  \sqrt{\lambda^2 + \epsilon} - \lambda = \lambda \left[ \left( 1 +
  \frac{\epsilon}{\lambda^2} \right)^{1/2} - 1 \right] \simeq
  \frac{\epsilon}{2 \lambda}.
\end{equation}
From here, we write
\begin{equation}
  \chi(\xpa) \simeq  \frac{1}{2 \lambda} \, e^{- \xpa \epsilon / 2
      \lambda}, 
\end{equation}
which allows to determine the correlation length $\xi_\parallel =
\epsilon^{-1}$. 

\subsection*{Static susceptibility in $d=2$}

In the case $d=2$, we have
\begin{equation}
  \chi(\xpa, \xpe) = \frac{1}{(2\pi)^2} \int_{-\infty}^{\infty} d k  \,
  e^{ik\xpe}\int_{-\infty}^{\infty} d q  \,
  \frac{e^{iq\xpa}}{k^2 + q^2 +  2 i\lambda q +\epsilon}.  
\end{equation}
Integrating by residues the $q$ variable, similarly as in the previous
subsection, we obtain
\begin{equation}
  \chi(\xpa, \xpe) = \frac{1}{2 \pi} e^{\lambda \xpa}
  \int_{0}^{\infty} d k \, \cos(k 
  \xpe) \frac{e^{-\xpa (\sqrt{k^2 + \lambda^2 +
  \epsilon})}}{\sqrt{k^2 
  + \lambda^2  + \epsilon}}.
\end{equation}
The last integral is duable \cite{formulitas}, giving a modified
Bessel function of order $0$:
\begin{equation}
  \chi(\xpa, \xpe) = \frac{1}{2 \pi} e^{\lambda \xpa} K_0 \left[
  (\lambda^2 + 
  \epsilon)^{1/2}  (\xpa^2 + \xpe^2)^{1/2} \right].
\end{equation}
In the limit $\xpa\gg\xpe\gg1$, the modified Bessel function $ K_0$
can be replaced by its asymptotic form $K_0(z) \simeq e^{-z}/ (2 \pi
z)^{1/2}$ \cite{formulitas}. Then, we have
\begin{equation}
  \chi(\xpa, \xpe) \simeq  \frac{1}{(2\pi)^{3/2}}  \frac{1}{(\lambda^2 
    + \epsilon)^{1/4}}   \frac{1}{(\xpa^2 + \xpe^2)^{1/4}}
  \exp\left( \lambda \xpa - [\lambda^2 + \epsilon]^{1/2} [\xpa^2 +
    \xpe^2]^{1/2} \right).
\end{equation}
In the double limit $\xpa\gg\xpe\gg1$ and $\epsilon\to0$, the argument
in the last exponential can be expanded
\begin{equation}
  \lambda \xpa - [\lambda^2 + \epsilon]^{1/2} [\xpa^2 + \xpe^2]^{1/2}
  \simeq \lambda \xpa - \left( \lambda + \frac{\epsilon}{2 \lambda}
  \right)  \left( \xpa + \frac{\xpe^2}{2 \xpa} \right) \simeq -
  \frac{\lambda \xpe^2}{2 \xpa} - \frac{\xpa \epsilon}{2 \lambda},
\end{equation}
where we have only kept terms linear in $\epsilon$ and $\xpe^2 /
\xpa$. 

The static susceptibility can be finally written 
\begin{equation}
  \chi(\xpa, \xpe) \simeq \frac{1}{(2\pi)^{3/2}}
  \frac{1}{\lambda^{1/2}} \xpa^{-1/2} 
  e^{ -\xpa \epsilon / 2 \lambda} e^{- \lambda \xpe^2 / 2 \xpa},
\end{equation}
from which we immediately read the correlation lengths $\xi_\parallel
= \epsilon^{-1}$ and $\xi_\perp = \epsilon^{-1/2}$.


\begin{thebibliography}{10}
\bibitem{notascaling}
It is obvious that, on the contrary, the exponents relating the deviation from
  the critical point can be different; i.e. cut-offs are determined by $L$ or
  $\epsilon$, which are different quantities.
\end{thebibliography}

\end{document}